# On the role of rotation-vibration coupling in the spectra of ozone isotopomers

Igor Gayday,[*] Alexander Teplukhin,[†] Brian K. Kendrick[†] and Dmitri Babikov[*1]

[*]*Department of Chemistry, Wehr Chemistry Building, Marquette University, Milwaukee, Wisconsin 53201-1881, USA*
[†]*Theoretical Division (T-1, MS B221), Los Alamos National Laboratory, Los Alamos, New Mexico 87545, USA*

## Abstract

A theoretical framework and computer code are developed for accurate calculations of coupled rotational-vibrational states in triatomic molecules using hyper-spherical coordinates and taking into account the Coriolis coupling effect. Concise final formulae are derived for construction of the Hamiltonian matrix using an efficient combination of the FBR and DVR methods with locally optimized basis sets and grids. First, the new code is tested by comparing its results with those of the APH3D program of Kendrick. Then, accurate calculations of the ro-vibrational spectra are carried out for doubly substituted symmetric ($^{18}O^{16}O^{18}O$) and asymmetric ($^{18}O^{18}O^{16}O$) ozone isotopomers for total angular momentum up to $J = 5$. Together with similar data recently reported for the singly substituted symmetric ($^{16}O^{18}O^{16}O$) and asymmetric ($^{16}O^{16}O^{18}O$) ozone isotopomers, these calculations quantify the role of the Coriolis coupling effect in the large mass-independent isotopic enrichment of ozone, observed in both laboratory experiments and the atmosphere of Earth. It is found that the Coriolis effect in ozone is relatively small, as evidenced by deviations of its rotational constants from the symmetric-top-rotor behavior, by the magnitudes of parity splittings (Λ-doubling), and by the ratios of ro-vibrational partition functions for asymmetric vs. symmetric ozone molecules. It is concluded that all of these characteristics are influenced by the isotopic masses as much as they are influenced by the overall symmetry of the molecule. It is therefore unlikely that the Coriolis coupling effect could be responsible for symmetry-driven mass-independent fractionation of oxygen isotopes in ozone.

[1] Author to whom all correspondence should be addressed; electronic mail: dmitri.babikov@mu.edu

## I. INTRODUCTION

Anomalously large and mass-independent enrichment of ozone molecules in heavy isotopes of oxygen ($^{17}$O and $^{18}$O *vs* $^{16}$O) has been a subject of numerous experimental[1,2,3,4,5,6,7,8] and theoretical[9,10,11,12,13,14,15,16] investigations during the last two decades. Experimental data[1] indicate that all symmetric ozone molecules are formed at very similar rates, regardless of their isotopic composition (*e.g.*, $^{16}$O$^{16}$O$^{16}$O, $^{16}$O$^{18}$O$^{16}$O, $^{18}$O$^{16}$O$^{18}$O). The same data also indicate that the asymmetric ozone molecules are formed faster on average than the symmetric ones, by as much as 16% of the rate[1], and that this effect is also independent of the number of isotopic substitutions (*e.g.*, the same for $^{16}$O$^{16}$O$^{18}$O and $^{18}$O$^{18}$O$^{16}$O). Clearly, this phenomenon is somehow driven by symmetry of the ozone molecule as a whole, rather than by the masses of its individual isotopes.[17]

Symmetry considerations on the global potential energy surface (PES) for the recombination reaction that forms ozone are surprisingly involved,[18] but if all the symmetry-related factors are taken into account properly, one concludes that the symmetry alone does not lead to any enrichments of the rare isotopes at all.[19] Recently it was proposed[20,21] that the Coriolis coupling effect may act more aggressively in the asymmetric ozone molecules, compared to the symmetric one, leading to more efficient flow of energy through the manifolds of the rotational-vibrational states, and thus increasing the rates of formation for these species. This hypothesis was tested by the classical trajectory simulations of rotational-vibrational excitations in the symmetric and asymmetric isotopomers of ozone.[20,21] Although no sufficient support was presented, the authors remained optimistic and concluded that: "*the symmetry effect of Coriolis coupling can appear in quantum mechanical analysis of the model.*"[20]

But, the quantum mechanical calculations of coupled rotational-vibrational motion are demanding both computationally and methodologically, and for this reason the Coriolis coupling terms were historically neglected in the calculations of the internal states of ozone. The earlier calculations of Schinke group[22] covered many important isotopologues and isotopomers of ozone but were restricted to the non-rotating ozone molecule only, $J = 0$. Several more recent calculations followed the same route,[13,23,24,25,26] being restricted only to $J = 0$ and $J = 1$ positive parity states, where just one rotational block occurs in the Hamiltonian matrix and thus no Coriolis coupling is possible. Rotationally excited ozone states were computed in several papers focused on the recombination reaction that forms ozone,[16,27,28,29] for a very broad range of rotational excitations up to $J\sim50$, but in all those cases the Coriolis coupling terms were neglected to ease calculations (using the symmetric-top rotor approximation, also known as $\Lambda$-conserving approximation, or the Coriolis-sudden approximation). The effect of ro-vibrational coupling can in principle be captured by semi-empirical methods of analysis, that use experimental information to adjust parameters of the effective Hamiltonian,[30,31] but this approach is not entirely general and its predictive power is limited (although it may be very accurate for a chosen part of spectrum of a given molecule).

Only very recently[32] the Coriolis coupling effect was included into accurate calculations of the rotational-vibrational states of ozone, up to $J = 5$. These calculations reported the spectra of symmetric $^{16}O^{18}O^{16}O$ and asymmetric $^{16}O^{16}O^{18}O$ isotopomers of singly substituted isotopologue of ozone, but the doubly substituted case was not covered. Since one of the goals is to reproduce by calculations the isotope effect that is both symmetry-driven and mass-independent (and eventually to explain this experimentally observed phenomenon), the doubly substituted case is absolutely necessary. One of the goals of this paper is to fill this gap and to provide the accurate theoretically derived rotational-vibrational spectra of the symmetric $^{18}O^{16}O^{18}O$ and asymmetric $^{18}O^{18}O^{16}O$ isotopomers of doubly substituted isotopologue of ozone. Interestingly, even at the $J = 0$ level, many authors choose to study only the singly substituted ozone,[23,24] and do not touch the doubly substituted case. The majority of information about the doubly substituted case comes from the experiment, and from the semi-empirical studies using the effective Hamiltonian approach.[30,31]

The computational methodology developed in this paper represents an important extension of our previous work,[25] that was done within the symmetric-rotor approximation. When this approximation is made, the asymmetric-top rotor contributions to the energy and the Coriolis coupling terms are both neglected. This decouples the calculations with different values of $\Lambda$ (projection of total angular momentum $J$ onto the z-axis), which then becomes a good quantum number. In the method developed here these couplings are taken into consideration, the relevant off-diagonal blocks of the Hamiltonian matrix are constructed and added at the last step of the calculations. This leads to an increase of the matrix-size by a factor of roughly $J$. The cost of matrix diagonalization typically grows as $J^3$, which raises the cost of the calculations dramatically. Here we outline a practical methodology that allows such calculations to be done in an efficient and accurate way. It is then applied to derive the spectra of doubly substituted ozone up to $J = 5$.

Together with the data already available for the singly substituted ozone, this gives us enough information to systematically analyze the magnitudes of the Coriolis coupling effect in symmetric and asymmetric isotopomers of singly and doubly substituted isotopologues of ozone: $^{16}O^{18}O^{16}O$, $^{16}O^{16}O^{18}O$, $^{18}O^{16}O^{18}O$ and $^{18}O^{18}O^{16}O$. Our results for these species show that the Coriolis coupling is affected by the number of isotopic substitutions (by the mass of ozone molecule) as much at it is affected by its symmetry. Unfortunately, in these data we could not identify any relevant trend that would clearly distinguish the asymmetric ozone molecules from the symmetric molecules and would be mass-independent at the same time. Therefore, the results presented in this work do not support the hypothesis[20] that the Coriolis coupling effect might be responsible for the mass-independent isotope fractionation phenomenon.

## II. THEORY

### A. Rotation-vibration Hamiltonian in hyper-spherical coordinates

The present theory is formulated in adiabatically adjusting principal-axes hyper-spherical (APH) coordinates.[18,33,34] Three APH coordinates $(\rho, \theta, \varphi)$ describe the shape of a triatomic system using the hyper-radial coordinate $\rho$, which represents the breathing motion, and two hyper-angles $\theta$ and $\varphi$, which correspond to the bending and asymmetric-stretching motions. Collectively, these are the vibrational degrees of freedom. Rotation of the molecule as a whole is described using the usual set of Euler angles $(\alpha, \beta, \gamma)$. The full rotation-vibration Hamiltonian operator in these coordinates includes the following terms:[25]

$$\hat{H} = \hat{T}_\rho + \hat{T}_\theta + \hat{T}_\varphi + V_{\text{pes}} + V_{\text{ext}} + \hat{T}_{\text{sym}} + \hat{T}_{\text{asym}} + \hat{T}_{\text{cor}}, \quad (1)$$

where the first three operators are associated with the kinetic energy along each vibrational degree of freedom and $V_{\text{ext}}$ is a potential-like term. Expressions for these operators are given by:[25]

$$\hat{T}_\rho = -\frac{\hbar^2}{2\mu} \frac{\partial^2}{\partial \rho^2}, \quad (2)$$

$$\hat{T}_\theta = -\frac{2\hbar^2}{\mu \rho^2} \frac{\partial^2}{\partial \theta^2}, \quad (3)$$

$$\hat{T}_\varphi = -\frac{2\hbar^2}{\mu \rho^2 \sin^2 \theta} \frac{\partial^2}{\partial \varphi^2}, \quad (4)$$

$$V_{\text{ext}} = -\frac{\hbar^2}{2\mu\rho^2}\left(\frac{1}{4} + \frac{4}{\sin^2 2\theta}\right). \tag{5}$$

The term $V_{\text{pes}}(\rho,\theta,\varphi)$ describes the electronic potential energy surface of the molecule under consideration. Expressions for the rotational operators $\hat{T}_{\text{sym}}$, $\hat{T}_{\text{asym}}$ and $\hat{T}_{\text{cor}}$ depend on whether the z-axis is chosen to be in the molecular plane or perpendicular to it, as well as whether the system is close to a prolate or to an oblate top. All ozone isotopomers considered here are close to a prolate top. In this case it is advantageous to place the z-axis in the plane of the molecule, as we discussed in detail recently.[35] Then:

$$\hat{T}_{\text{sym}} = \frac{A+B}{2}\hat{J}^2 + \left(C - \frac{A+B}{2}\right)\hat{J}_z^2, \tag{6}$$

$$\hat{T}_{\text{asym}} = \frac{A-B}{2}(\hat{J}_x^2 - \hat{J}_y^2), \tag{7}$$

$$\hat{T}_{\text{cor}} = 4B\cos\theta\left(i\hbar\frac{\partial}{\partial\varphi}\right)\hat{J}_y, \tag{8}$$

where the rotational constants $A$, $B$ and $C$ are given by the following expressions:

$$A^{-1} = \mu\rho^2(1 + \sin\theta), \tag{9}$$

$$B^{-1} = 2\mu\rho^2 \sin^2\theta, \tag{10}$$

$$C^{-1} = \mu\rho^2(1 - \sin\theta). \tag{11}$$

The volume element for computing matrix elements of this operator is given by:

$$d^6 v = d\rho\, d\theta\, d\varphi\, d\alpha\, \sin(\beta)\, d\beta\, d\gamma. \tag{12}$$

**B. Wave function and Hamiltonian matrix**

The full-dimensional ro-vibrational wave functions (6D) can be represented by an expansion over the rotational components $\tilde{D}_\Lambda(\alpha,\beta,\gamma)$, where the vibrational components $\Psi_\Lambda^k(\rho,\theta,\varphi)$ play the role of expansion coefficients, namely:

$$F^k(\rho,\theta,\varphi,\alpha,\beta,\gamma) = \sum_{\Lambda=0,1}^{J} \Psi_\Lambda^k(\rho,\theta,\varphi)\tilde{D}_\Lambda(\alpha,\beta,\gamma), \tag{13}$$

$$\hat{H}F^k(\rho,\theta,\varphi,\alpha,\beta,\gamma) = \varepsilon^k F^k(\rho,\theta,\varphi,\alpha,\beta,\gamma). \tag{14}$$

The rotational basis functions $\tilde{D}_\Lambda(\alpha,\beta,\gamma)$ are taken in the form of the modified normalized Wigner D-functions of two parities ($p=0$ and $p=1$):

$$\widetilde{D}_{\Lambda M}^{Jp} = \sqrt{\frac{2J+1}{16\pi^2(1+\delta_{\Lambda 0})}} \left[ D_{\Lambda M}^{J}(\alpha,\beta,\gamma) + (-1)^{J+\Lambda+p} D_{-\Lambda M}^{J}(\alpha,\beta,\gamma) \right]. \tag{15}$$

The starting value of $\Lambda$ in Eq. (13) is 0 if $J + p$ is even or 1 otherwise. Different values of $J$, $M$ and $p$ are not coupled with each other and the corresponding calculations can be carried out independently. Since the values of $J$, $M$ and $p$ stay constant within each calculation, their indexes are assumed implicit in Eq. (13) and are omitted further in the text, for clarity.

Each vibrational component $\Psi_\Lambda^k(\rho,\theta,\varphi)$ in Eq. (13) is individually expanded as:

$$\Psi_\Lambda^k(\rho,\theta,\varphi) = \sum_n^N \sum_j^{S_{\Lambda n}} c_{\Lambda nj}^k h_n(\rho) X_{\Lambda n}^j(\theta,\varphi), \tag{16}$$

where $h_n(\rho)$ are the DVR basis functions for the hyper-radius, and $X_{\Lambda n}^j(\theta,\varphi)$ represents a locally optimized basis set for the hyper-angles (discussed in the next subsection). In this basis, the matrix elements of the Hamiltonian operator of Eq. (1) are given by:

$$\begin{aligned}
\langle h_n X_{\Lambda n}^j \widetilde{D}_\Lambda | \hat{H} | h_{n'} X_{\Lambda' n'}^{j'} \widetilde{D}_{\Lambda'} \rangle &= \langle h_n | \hat{T}_\rho | h_{n'} \rangle \langle X_{\Lambda n}^j | X_{\Lambda' n'}^{j'} \rangle \langle \widetilde{D}_\Lambda | \widetilde{D}_{\Lambda'} \rangle \\
&+ \langle h_n X_{\Lambda n}^j | \hat{T}_\theta + \hat{T}_\varphi + V_{\text{pes}} + V_{\text{ext}} | h_{n'} X_{\Lambda' n'}^{j'} \rangle \langle \widetilde{D}_\Lambda | \widetilde{D}_{\Lambda'} \rangle \\
&+ \langle h_n X_{\Lambda n}^j \widetilde{D}_\Lambda | \hat{T}_{\text{sym}} + \hat{T}_{\text{asym}} + \hat{T}_{\text{cor}} | h_{n'} X_{\Lambda' n'}^{j'} \widetilde{D}_{\Lambda'} \rangle \\
&= \langle h_n | \hat{T}_\rho | h_{n'} \rangle \langle X_{\Lambda n}^j | X_{\Lambda n'}^{j'} \rangle \tilde{\delta}_{\Lambda \Lambda'} \\
&+ \langle X_{\Lambda n}^j | \hat{T}_\theta^n + \hat{T}_\varphi^n + V_{\text{pes}}^n + V_{\text{ext}}^n | X_{\Lambda n}^{j'} \rangle \delta_{nn'} \tilde{\delta}_{\Lambda \Lambda'} \\
&+ \langle h_n X_{\Lambda n}^j \widetilde{D}_\Lambda | \hat{T}_{\text{sym}} + \hat{T}_{\text{asym}} + \hat{T}_{\text{cor}} | h_{n'} X_{\Lambda' n'}^{j'} \widetilde{D}_{\Lambda'} \rangle. \tag{17}
\end{aligned}$$

Here we introduced the following notation:

$$\tilde{\delta}_{\Lambda \Lambda'} = \begin{cases} \delta_{\Lambda \Lambda'} & \text{if } \Lambda, \Lambda' \neq 0, \\ \delta_{(-1)^{J+p},1} & \text{if } \Lambda, \Lambda' = 0. \end{cases} \tag{18}$$

where $\delta$ is the usual Kronecker symbol. Since the DVR basis functions $h_n(\rho)$ are non-zero only at $\rho = \rho_n$ (see Appendix A), it was also convenient to introduce, in the second term of Eq. (17), the operators $\hat{T}_\theta^n$ and $\hat{T}_\varphi^n$ and the functions $V_{\text{pes}}^n$ and $V_{\text{ext}}^n$ specific to the $n$-th point of the $\rho$-grid.

The last term in Eq. (17) is responsible for the rotation-vibration interaction. In a previous work we derived expressions for the matrix elements of the operators $\hat{T}_{\text{sym}}$, $\hat{T}_{\text{asym}}$ and $\hat{T}_{\text{cor}}$. Building upon that work, here we can write:[35]

$$\langle h_n X_{\Lambda n}^{j} \widetilde{D}_\Lambda | \hat{T}_{\text{sym}} | h_{n'} X_{\Lambda' n'}^{j'} \widetilde{D}_{\Lambda'} \rangle = \langle h_n X_{\Lambda n}^{j} | V_{\text{rot}}^{\Lambda} | h_{n'} X_{\Lambda n'}^{j'} \rangle \widetilde{\delta}_{\Lambda\Lambda'}$$
$$= \langle X_{\Lambda n}^{j} | V_{\text{rot}}^{\Lambda n} | X_{\Lambda n}^{j'} \rangle \delta_{nn'} \widetilde{\delta}_{\Lambda\Lambda'}, \tag{19}$$

$$\langle h_n X_{\Lambda n}^{j} \widetilde{D}_\Lambda | \hat{T}_{\text{asym}} | h_{n'} X_{\Lambda' n'}^{j'} \widetilde{D}_{\Lambda'} \rangle = \frac{\hbar^2}{4} \langle X_{\Lambda n}^{j} | A_n - B_n | X_{\Lambda' n}^{j'} \rangle \delta_{nn'} U_{\Lambda\Lambda'}, \tag{20}$$

$$\langle h_n X_{\Lambda n}^{j} \widetilde{D}_\Lambda | \hat{T}_{cor} | h_{n'} X_{\Lambda' n'}^{j'} \widetilde{D}_{\Lambda'} \rangle = 2\hbar^2 \langle X_{\Lambda n}^{j} | B_n \cos\theta \frac{d}{d\varphi} | X_{\Lambda' n}^{j'} \rangle \delta_{nn'} W_{\Lambda\Lambda'}. \tag{21}$$

Note that in Eq. (19) the rotational potential of a symmetric-top rotor was introduced:

$$V_{\text{rot}}^{\Lambda}(\rho, \theta) = \hbar^2 \left( J(J+1) \frac{A+B}{2} + \Lambda^2 \left( C - \frac{A+B}{2} \right) \right). \tag{22}$$

The rotational matrixes $U_{\Lambda\Lambda'}$ and $W_{\Lambda\Lambda'}$ in Eqs. (20) and (21) are analytic (see Appendix B). The remaining vibrational integrals in Eqs. (19-21) are over the hyper-angles $\theta$ and $\varphi$, and should be computed numerically.

The structure of the Hamiltonian matrix in Eq. (17) can be greatly simplified if the vibrational basis functions $X_{\Lambda n}^{j}(\theta, \varphi)$ are chosen to be the eigenfunctions of the 2D operator in hyper-angles $\theta$ and $\varphi$:

$$\hat{H}_{2D}^{\Lambda n} = \hat{T}_\theta^n + \hat{T}_\varphi^n + V_{\text{pes}}^n + V_{\text{ext}}^n + V_{\text{rot}}^{\Lambda n}, \tag{23}$$

*i.e.*:

$$\hat{H}_{2D}^{\Lambda n} X_{\Lambda n}^{j}(\theta, \varphi) = \varepsilon_{\Lambda n}^{j} X_{\Lambda n}^{j}(\theta, \varphi). \tag{24}$$

Note that for each $\rho_n$ this operator includes the rotational potential $V_{\text{rot}}^{\Lambda n}$ of a symmetric top rotor, just like in Eq. (22), but with $\rho = \rho_n$. Since each set of $X_{\Lambda n}^{j}(\theta, \varphi)$ is orthonormal, Eq. (17) transforms into the following final expression for the ro-vibrational Hamiltonian matrix:

$$\langle h_n X_{\Lambda n}^{j} \widetilde{D}_\Lambda | \hat{H} | h_{n'} X_{\Lambda' n'}^{j'} \widetilde{D}_{\Lambda'} \rangle = \widetilde{\delta}_{\Lambda\Lambda'} \langle h_n | \hat{T}_\rho | h_{n'} \rangle \langle X_{\Lambda n}^{j} | X_{\Lambda n'}^{j'} \rangle$$
$$+ \widetilde{\delta}_{\Lambda\Lambda'} \delta_{nn'} \delta_{jj'} \varepsilon_{\Lambda n}^{j}$$
$$+ \frac{\hbar^2}{4} U_{\Lambda\Lambda'} \delta_{nn'} \langle X_{\Lambda n}^{j} | A_n - B_n | X_{\Lambda' n}^{j'} \rangle$$
$$+ 2\hbar^2 W_{\Lambda\Lambda'} \delta_{nn'} \langle X_{\Lambda n}^{j} | B_n \cos\theta \frac{d}{d\varphi} | X_{\Lambda' n}^{j'} \rangle. \tag{25}$$

The last two terms of this expression are responsible for the ro-vibrational coupling and correspond to the asymmetric top rotor energy and the Coriolis coupling effect, respectively. The rotational block structure of this matrix was discussed in detail in recent work.[35] The structure of its vibrational blocks was discussed earlier in Ref. 25. This matrix is diagonalized numerically to determine the eigenvalues of the ro-vibrational energy in Eq. (14).

### C. Sequential diagonalization-truncation (SDT)

In order to determine a suitable set of 2D functions $X_{\Lambda n}^{j}(\theta, \varphi)$ the hierarchy of expansions is continued. Namely, for each point $n$ of $\rho$-grid and for each $\Lambda$ the following expansion is constructed:

$$X_{\Lambda n}^{j}(\theta, \varphi) = \sum_{l}^{L} \sum_{i}^{S_{\Lambda nl}} b_{\Lambda nli}^{j} g_l(\theta) \Phi_{\Lambda nl}^{i}(\varphi), \tag{26}$$

where $g_l(\theta)$ is a set of DVR basis functions for the hyper-angle $\theta$ (defined in Appendix A), while $\Phi_{\Lambda nl}^{i}(\varphi)$ is a locally-optimal basis set of functions for the remaining hyper-angle $\varphi$. Matrix elements of $\widehat{H}_{2D}^{\Lambda n}$ in this basis are given by:

$$\langle g_l \Phi_{\Lambda nl}^{i} | \widehat{H}_{2D}^{\Lambda n} | g_{l'} \Phi_{\Lambda nl'}^{i'} \rangle = \langle g_l | \widehat{T}_{\theta}^{n} | g_{l'} \rangle \langle \Phi_{\Lambda nl}^{i} | \Phi_{\Lambda nl'}^{i'} \rangle \tag{27}$$
$$+ \langle \Phi_{\Lambda nl}^{i} | \widehat{T}_{\varphi}^{nl} + V_{\text{pes}}^{nl} + V_{\text{ext}}^{nl} + V_{\text{rot}}^{\Lambda nl} | \Phi_{\Lambda nl}^{i'} \rangle \delta_{ll'}.$$

Once again, $g_l(\theta)$ is non-zero only at $\theta = \theta_l$, so one can set $\theta = \theta_l$ in the operator $\widehat{T}_{\varphi}^{n}$ and in the functions $V_{\text{pes}}^{n}$, $V_{\text{ext}}^{n}$ and $V_{\text{rot}}^{\Lambda n}$, by introducing in Eq. (27) their versions labelled by $l$. Again, the structure of this matrix is simplified by choosing $\Phi_{\Lambda nl}^{i}(\varphi)$ to be the eigenfunctions of the 1D operator in hyper-angle $\varphi$:

$$\widehat{H}_{1D}^{\Lambda nl} = \widehat{T}_{\varphi}^{nl} + V_{\text{pes}}^{nl} + V_{\text{ext}}^{nl} + V_{\text{rot}}^{\Lambda nl}, \tag{28}$$

*i.e.*:

$$\widehat{H}_{1D}^{\Lambda nl} \Phi_{\Lambda nl}^{i}(\varphi) = \varepsilon_{\Lambda nl}^{i} \Phi_{\Lambda nl}^{i}(\varphi). \tag{29}$$

Since each of these sets is orthonormal, we obtain:

$$\langle g_l \Phi_{\Lambda nl}^{i} | \widehat{H}_{2D}^{\Lambda n} | g_{l'} \Phi_{\Lambda nl'}^{i'} \rangle = \langle \Phi_{\Lambda nl}^{i} | \Phi_{\Lambda nl'}^{i'} \rangle \langle g_l | \widehat{T}_{\theta}^{n} | g_{l'} \rangle + \varepsilon_{\Lambda nl}^{i} \delta_{ii'} \delta_{ll'}. \tag{30}$$

And lastly, the locally optimal sets of 1D functions $\Phi_{\Lambda nl}^{i}(\varphi)$ are determined using the FBR for hyper-angle $\varphi$:

$$\Phi_{\Lambda nl}^{i}(\varphi) = \sum_{m}^{M} a_{\Lambda nlm}^{i} f_m(\varphi). \tag{31}$$

The basis functions $f_m(\varphi)$ are symmetry-adapted (sine and cosine functions, see Appendix A), since in the APH coordinates the symmetry of vibrational wavefunctions is defined with respect to this variable. The matrix elements of $\widehat{H}_{1D}^{\Lambda nl}$ in this basis are:

$$\langle f_m|\widehat{H}_{1D}^{\Lambda nl}|f_{m'}\rangle = \langle f_m|V_{pes}^{nl}|f_{m'}\rangle - m^2 \delta_{mm'} + (V_{ext}^{nl} + V_{rot}^{\Lambda nl})\delta_{mm'} . \tag{32}$$

Here, the matrix elements of the PES are computed numerically (using a large 1D quadrature in $\varphi$), while matrix elements of the kinetic energy operator in $\varphi$ are analytic. The last term in this formula is just a constant energy shift of each individual 1D problem, since $V_{ext}$ and $V_{rot}^{\Lambda}$ do not depend on $\varphi$, according to Eqs. (5) and (22), with definitions of Eqs. (9-11).

Practical implementation of this approach proceeds in the reverse order, starting from 1D and going to 6D. The first step is the calculation of eigenvalues $\varepsilon_{\Lambda nl}^i$ and eigenvectors $a_{\Lambda nlm}^i$ for each of the $\Lambda \times n \times l$ one-dimensional operators $\widehat{H}_{1D}^{\Lambda nl}$, by diagonalization of the corresponding matrixes given by Eq. (32). Before proceeding to the next step, this set of 1D solutions is truncated based on their energy, to keep only the solutions with $\varepsilon_{\Lambda nl}^i < E_{cut}$, where $E_{cut}$ is a convergence parameter that depends on the system and the energy span of the spectrum (here $E_{cut} = 6000$ cm$^{-1}$ above the dissociation threshold of O$_3$ was used). The retained solutions represent the locally optimal 1D-basis sets $\Phi_{\Lambda nl}^i(\varphi)$ for Eq. (26).

The second step is the calculation of eigenvalues $\varepsilon_{\Lambda n}^j$ and eigenvectors $b_{\Lambda nli}^j$ for each of the $\Lambda \times n$ two-dimensional operators $\widehat{H}_{2D}^{\Lambda n}$, by diagonalization of the corresponding matrixes given by Eq. (27). Again, before proceeding to the next step, this set of 2D solutions is truncated using the same energy criterion $\varepsilon_{\Lambda n}^j < E_{cut}$ to determine the locally-optimized 2D-basis sets $X_{\Lambda n}^j(\theta, \varphi)$ for Eq. (16). It should be emphasized that this method adjusts basis sets locally to the shape of the PES, but also takes into account the level of rotational excitation of the system (determined by the values of $J$ and $\Lambda$), since the rotational potential $V_{rot}^{\Lambda nl}$ is introduced at the very beginning, in Eq. (28).

At the final third step a set of three-dimensional vibrational eigenvectors $c_{\Lambda nj}^k$ is obtained by diagonalizing the Hamiltonian matrix of Eq. (25), which also takes into account the rotational-vibrational couplings (the asymmetric top rotor terms and the Coriolis coupling terms, including the effect of parity $p$). This gives the spectrum of coupled rotational-vibrational eigenstates of the system, $\varepsilon^k$, and the overall 6D ro-vibrational wave function, expressed by combination of Eqs. (31), (26), (16) and (13), as follows:

$$F^k(\rho,\theta,\phi,\alpha,\beta,\gamma) = \sum_{\Lambda=0,1}^{J}\sum_{n}^{N}\sum_{j}^{S_{\Lambda n}}\sum_{l}^{L}\sum_{i}^{S_{\Lambda nl}}\sum_{m}^{M} c_{\Lambda nj}^k b_{\Lambda nli}^j a_{\Lambda nlm}^i h_n(\rho) g_l(\theta) f_m(\varphi) \widetilde{D}_\Lambda(\alpha,\beta,\gamma). \tag{33}$$

Such sequential addition of the vibrational degrees of freedom, with truncation of solutions between the steps, is known as the Sequential Diagonalization Truncation (SDT) method.[36] It reduces the size of the Hamiltonian matrix, and in this way achieves a significant computational advantage, in comparison with a brute-force application of the multi-dimensional basis sets represented by a direct-product of generic DVR of FBR functions.[37]

**D. Calculation of the vibrational matrix elements**

Calculation of the matrix elements of the kinetic energy operators in the DVR basis, $\langle h_n|\hat{T}_\rho|h_{n'}\rangle$ in Eq. (25) and $\langle g_l|\hat{T}_\theta^n|g_{l'}\rangle$ in Eq. (30), is discussed in Appendix A. Calculation of the matrix elements for the overlap matrixes at the 1D-level, 2D-level and the 3D-level is carried out analytically using their expansion coefficients, as follows:

$$\begin{aligned}\langle \Phi_{\Lambda nl}^i | \Phi_{\Lambda nl'}^{i'} \rangle &= \langle \sum_m^M a_{\Lambda nlm}^i f_m | \sum_{m'}^M a_{\Lambda nl'm'}^{i'} f_{m'} \rangle = \sum_m^M \sum_{m'}^M a_{\Lambda nlm}^i a_{\Lambda nl'm'}^{i'} \langle f_m | f_{m'} \rangle \\ &= \sum_m^M a_{\Lambda nlm}^i a_{\Lambda nl'm}^{i'}, \end{aligned} \tag{34}$$

$$\begin{aligned}\langle X_{\Lambda n}^j | X_{\Lambda n'}^{j'} \rangle &= \langle \sum_l^L \sum_i^{S_{\Lambda nl}} b_{\Lambda nli}^j g_l \Phi_{\Lambda nl}^i | \sum_{l'}^L \sum_{i'}^{S_{\Lambda n'l'}} b_{\Lambda n'l'i'}^{j'} g_{l'} \Phi_{\Lambda n'l'}^{i'} \rangle \\ &= \sum_l^L \sum_i^{S_{\Lambda nl}} \sum_{l'}^L \sum_{i'}^{S_{\Lambda nl}} b_{\Lambda nli}^j b_{\Lambda n'l'i'}^{j'} \langle g_l | g_{l'} \rangle \langle \Phi_{\Lambda nl}^i | \Phi_{\Lambda n'l'}^{i'} \rangle \\ &= \sum_l^L \sum_m^M \left( \sum_i^{S_{\Lambda nl}} b_{\Lambda nli}^j a_{\Lambda nlm}^i \right) \left( \sum_{i'}^{S_{\Lambda n'l}} b_{\Lambda n'li'}^{j'} a_{\Lambda n'lm}^{i'} \right), \end{aligned} \tag{35}$$

$$\langle F^k | F^{k'} \rangle$$
$$= \langle \sum_{\Lambda,n,j,l,i,m} c^k_{\Lambda n j} b^j_{\Lambda n l i} a^i_{\Lambda n l m} h_n g_l f_m \widetilde{D}_\Lambda | \sum_{\Lambda',n',j',l',i',m'} c^{k'}_{\Lambda' n' j'} b^{j'}_{\Lambda' n' l' i'} a^{i'}_{\Lambda' n' l' m'} h_{n'} g_{l'} f_{m'} \widetilde{D}_{\Lambda'} \rangle$$
$$= \sum_{\Lambda=0,1}^{J} \sum_n^N \sum_l^L \sum_m^M \left( \sum_j^{S_{\Lambda n}} c^{k*}_{\Lambda n j} \sum_i^{S_{\Lambda n l}} b^j_{\Lambda n l i} a^i_{\Lambda n l m} \right) \left( \sum_{j'}^{S_{\Lambda n}} c^{k'}_{\Lambda n j'} \sum_{i'}^{S_{\Lambda n l}} b^{j'}_{\Lambda n l i'} a^{i'}_{\Lambda n l m} \right) = \delta_{kk'} .$$
(36)

In Eq. (36) the outer sum starts at $\Lambda = 0$ when $J + p$ is even, but it starts at $\Lambda = 1$ when $J + p$ is odd. Also note that Eq. (34) gives $\delta_{ii'}$ in the case of $l = l'$ due to orthonormality of the functions $\Phi^i_{\Lambda n l}(\varphi)$. Similar, Eq. (35) in the case of $n = n'$ gives $\delta_{jj'}$ due to orthonormality of the functions $X^j_{\Lambda n}(\theta, \varphi)$.

The matrix elements of the asymmetric top rotor kinetic energy term in Eq. (25) are computed analytically, similar to Eq. (35):

$$\langle X^j_{\Lambda n} | A_n - B_n | X^{j'}_{\Lambda' n} \rangle$$
$$= \langle \sum_l^L \sum_i^{S_{\Lambda n l}} b^j_{\Lambda n l i} g_l \Phi^i_{\Lambda n l} | A_n - B_n | \sum_{l'}^L \sum_{i'}^{S_{\Lambda' n l'}} b^{j'}_{\Lambda' n l' i'} g_{l'} \Phi^{i'}_{\Lambda' n l'} \rangle$$
$$= \sum_l^L (A_{nl} - B_{nl}) \sum_m^M \left( \sum_i^{S_{\Lambda n l}} b^j_{\Lambda n l i} a^i_{\Lambda n l m} \right) \left( \sum_{i'}^{S_{\Lambda' n l}} b^{j'}_{\Lambda' n l i'} a^{i'}_{\Lambda' n l m} \right) .$$
(37)

The matrix elements of the Coriolis coupling in Eq. (25) are also computed analytically:

$$\langle X^j_{\Lambda n} | B_n \cos\theta \frac{d}{d\varphi} | X^{j'}_{\Lambda' n} \rangle$$
$$= \langle \sum_l^L \sum_i^{S_{\Lambda n l}} \sum_m^M b^j_{\Lambda n l i} a^i_{\Lambda n l m} g_l f_m | B_n \cos\theta \frac{d}{d\varphi} | \sum_{l'}^L \sum_{i'}^{S_{\Lambda' n l'}} \sum_{m'}^M b^{j'}_{\Lambda' n l' i'} a^{i'}_{\Lambda' n l' m'} g_{l'} f_{m'} \rangle$$
$$= \sum_{l,i,m} \sum_{l',i',m'} b^j_{\Lambda n l i} a^i_{\Lambda n l m} b^{j'}_{\Lambda' n l' i'} a^{i'}_{\Lambda' n l' m'} \langle g_l | B_n \cos\theta | g_{l'} \rangle \langle f_m | \frac{d}{d\varphi} | f_{m'} \rangle$$
$$= (-1)^{\Lambda+s} \sum_l^L B_{nl} \cos\theta_l \sum_m^M m \left( \sum_i^{S_{\Lambda n l}} b^j_{\Lambda n l i} a^i_{\Lambda n l m} \right) \left( \sum_{i'}^{S_{\Lambda' n l}} b^{j'}_{\Lambda' n l i'} a^{i'}_{\Lambda' n l m} \right) .$$
(38)

Computation of $\langle f_m | \frac{d}{d\varphi} | f_{m'} \rangle$ is described in Appendix A. Combining all the results above, the coupled rotational-vibrational Hamiltonian matrix of Eq. (25) is expressed through the expansion coefficients of 1D and 2D functions as follows:

$$\langle h_n X^j_{\Lambda n} \widetilde{D}_\Lambda | \hat{H} | h_{n'} X^{j'}_{\Lambda' n'} \widetilde{D}_{\Lambda'} \rangle$$

$$= \tilde{\delta}_{\Lambda\Lambda'} \left( \langle h_n | \hat{T}_\rho | h_{n'} \rangle \sum_l^L \sum_m^M \left( \sum_i^{S_{\Lambda nl}} b^j_{\Lambda nli} a^i_{\Lambda nlm} \right) \left( \sum_{i'}^{S_{\Lambda n'l}} b^{j'}_{\Lambda n'li'} a^{i'}_{\Lambda n'lm} \right) + \delta_{nn'} \delta_{jj'} \varepsilon^j_{\Lambda n} \right)$$

$$+ \frac{\hbar^2}{4} U_{\Lambda\Lambda'} \delta_{nn'} \sum_l^L (A_{nl} - B_{nl}) \sum_m^M \left( \sum_i^{S_{\Lambda nl}} b^j_{\Lambda nli} a^i_{\Lambda nlm} \right) \left( \sum_{i'}^{S_{\Lambda' nl}} b^{j'}_{\Lambda' nli'} a^{i'}_{\Lambda' nlm} \right)$$

$$+ (-1)^{\Lambda+s} 2\hbar^2 W_{\Lambda\Lambda'} \delta_{nn'} \sum_l^L B_{nl} \cos\theta_l \sum_m^M m \left( \sum_i^{S_{\Lambda nl}} b^j_{\Lambda nli} a^i_{\Lambda nlm} \right) \left( \sum_{i'}^{S_{\Lambda' nl}} b^{j'}_{\Lambda' nli'} a^{i'}_{\Lambda' nlm} \right). \quad (39)$$

### E. Assignments of the ro-vibrational states

In contrast to the symmetric top rotor approximation (where the overall Hamiltonian does not have couplings between different values of Λ, and thus each wave function can be characterized by one value of Λ), the fully coupled ro-vibrational wave functions $F^k$ have a probability distribution over multiple values of Λ. For each Λ this probability is given by the respective term of the outer sum in Eq. (36), so we can write:

$$\langle F^k | F^{k'} \rangle = \sum_{\Lambda=0,1}^J P^k_\Lambda = \delta_{kk'}. \quad (40)$$

For all ro-vibrational states calculated in this work we computed the values of $P_\Lambda$ and found that the majority of states are still localized in one dominant value of Λ, so this value can still be used to label the ro-vibrational states, just like in the case of the symmetric top rotor approximation. We also saw that when the energies of two states are close to each other, they may display a mixture of several values of Λ, but such cases are relatively rare. Namely, among all the states computed in this work (7200 states overall) we found only one pair of energetically close states where the weights of two largest Λ-components were in the ratio close to 50/50. We also saw two examples when the two largest Λ-components gave the ratio of about 80/20. For all other states the weight of the second largest value of Λ was below 5%.

The global PES of $O_3$ features three energetically equivalent wells, one for symmetric isotopomers and two for asymmetric isotopomers of ozone, for example $^{16}O^{18}O^{16}O$ vs $^{16}O^{16}O^{18}O$ (in the singly substituted case) or $^{18}O^{16}O^{18}O$ vs $^{18}O^{18}O^{16}O$ (in the doubly substituted case). Figure 1 represents a map of the PES which shows that the wells can be easily separated using the value of hyper-angle $\varphi$. The wells of asymmetric isotopomers are centered at $\varphi = \pm \pi/3$, whereas the well of the symmetric isotopomer is centered at $\varphi = \pi$. Therefore, it is convenient to define a formal operator that acts on the basis functions $f_m(\varphi)$ by "cutting out" the part of wave function that corresponds to the symmetric isotopomer:

$$\hat{P}_{sym} f_m(\varphi) = \begin{cases} f_m(\varphi) \text{ for } \varphi \in \left[\frac{2\pi}{3}, \frac{4\pi}{3}\right], \\ 0 \text{ for } \varphi \notin \left[\frac{2\pi}{3}, \frac{4\pi}{3}\right]. \end{cases} \quad (41)$$

With this operator, the probability that a given state $F^k$ is a state of a symmetric molecule is given by $P^k_{sym} = \langle F^k | \hat{P}_{sym} | F^k \rangle$. Since we have only two kinds of isotopomers, either symmetric or asymmetric, the probability that a given state is a state of an asymmetric molecule can be calculated simply as $P^k_{asym} = 1 - P^k_{sym}$. Expressing the value of this integral in terms of expansion coefficients of the wave function, we obtain:

$$\langle F^k | \hat{P}_{sym} | F^k \rangle = \sum_{\Lambda,n,l,j,i,m,j',i',m'} c^{k*}_{\Lambda n j} b^{j}_{\Lambda n l i} a^{i}_{\Lambda n l m} c^{k}_{\Lambda n j'} b^{j'}_{\Lambda n l i'} a^{i'}_{\Lambda n l m'} \langle f_m | \hat{P}_{sym} | f_{m'} \rangle$$

$$= \sum_{\Lambda=0,1}^{J} \sum_{n}^{N} \sum_{l}^{L} \sum_{m}^{M} \sum_{m'}^{M} \langle f_m | \hat{P}_{sym} | f_{m'} \rangle \left( \sum_{j}^{S_{\Lambda n}} c^{k*}_{\Lambda n j} \sum_{i}^{S_{\Lambda n l}} b^{j}_{\Lambda n l i} a^{i}_{\Lambda n l m} \right) \left( \sum_{j'}^{S_{\Lambda n}} c^{k}_{\Lambda n j'} \sum_{i'}^{S_{\Lambda n l}} b^{j'}_{\Lambda n l i'} a^{i'}_{\Lambda n l m'} \right). \quad (42)$$

Here, in contrast to Eq. (36), we cannot eliminate the sum over $m'$. The integral $\langle f_m | \hat{P}_{sym} | f_{m'} \rangle$ over $\varphi$ is computed analytically as shown in Appendix A.

In this way we computed the values of $P_{sym}$ and $P_{asym}$ for every ro-vibrational state reported in this work and found that the states of symmetric and asymmetric isotopomers never mix. Namely, if the value of $P_{sym}$ is on the order of one, then the value of $P_{asym}$ is on the order of $10^{-13}$, and vice versa. Thus, the assignment of isotopomers to the states can be confidently used, at least in this part of spectrum of ozone.

Finally, the coupled rotational-vibrational states are labelled by their symmetry ($A_1$, $A_2$, $B_1$ and $B_2$) as it was discussed in detail elsewhere,[35] and is briefly reviewed in Appendix A.

## III. RESULTS AND DISCUSSION

### A. Benchmark tests

In order to verify that all pieces of our theory are correct, and the new code we wrote works as expected, we carried out a set of benchmark studies using two different computer programs. In addition to the code developed in this work, we used the code developed earlier by Kendrick.[38] The code of Kendrick also uses the APH coordinates, but it is different in many respects. First of all, it starts with a general Fourier basis $e^{\pm im\varphi}$ for the hyper-angle angle $\varphi$, and the vibrational states of two symmetries are projected out only at the 2D level. In contrast, in the code developed here the two symmetries are treated separately from the very beginning, by employing the real-valued basis sets of either $\sin(m\varphi)$ or $\cos(m\varphi)$ functions. Second, the code of Kendrick uses an FBR of polynomials for the hyper-angle $\theta$, while here a simple DVR grid is used. Third, the code of Kendrick solves the coupled-channel equations for hyper-radius $\rho$ using the method of Numerov, while here we implement one more level of truncation and then build and diagonalize the Hamiltonian matrix for the vibrational 3D problem, using a DVR grid in $\rho$ optimized to the shape of the PES.[39,25] Finally, for the description of rotation the code of Kendrick uses the z-axis perpendicular to the plane of the molecule and includes the Coriolis terms from the beginning, while in the new code developed here the z-axis is placed in the molecular plane and the Coriolis couplings are taken into account only at the last step of calculations.

Rotational-vibrational states of both parities ($p = 0$ and $p = 1$) were computed using these two codes for $J = 3$ of the doubly-substituted ozone, both symmetric $^{18}O^{16}O^{18}O$ and asymmetric $^{18}O^{18}O^{16}O$ isotopomers, up to the energy of about 5200 cm$^{-1}$ above the bottom of the well, which is about 4800 cm$^{-1}$ below the dissociation threshold (roughly, one hundred vibrational states per each value of $\Lambda = 0, 1, 2, 3$). The absolute values of energy differences between the corresponding states computed with the two codes are presented in Figure 2. As one can see, the majority of the states agree to within $10^{-3}$ cm$^{-1}$ or better, reaching the difference of about 0.05 cm$^{-1}$ in the worst case at the high energy part of the spectrum. The overall agreement between the results of the two codes allows us to conclude that the ro-vibrational wave functions and their energies reported in this work are computed correctly.

## B. Ro-vibrational spectra of $^{18}O^{16}O^{18}O$ and $^{18}O^{18}O^{16}O$

Using our new code we computed the ro-vibrational states of doubly-substituted ozone molecules, $^{18}O^{16}O^{18}O$ and $^{18}O^{18}O^{16}O$, for all values of rotational excitations up to $J = 5$ using an optimized grid of 90 DVR functions $h_n(\rho)$ in the range of $3.4 \leq \rho \leq 6.1$ Bohr, 130 DVR functions $g_l(\theta)$ in the range $0.43 \leq \theta \leq 1.56$ radians, and 100 FBR functions $f_m(\varphi)$ of each symmetry. The truncation energy threshold for the SDT was set to $E_{\text{cut}} = 6000$ cm$^{-1}$. With these parameters, the size of the vibrational 3D Hamiltonian matrix was about 17500 rows/columns per each rotational Λ-block. We checked and found that variation of each convergence parameter by ±20% changes the state energies by no more than $10^{-3}$ cm$^{-1}$, which is our targeted accuracy here.

All energies computed for the doubly substituted ozone, 3600 states total, are reported in Section C of *Supplementary Information*, where the states are assigned by $J$, parity, dominant Λ, symmetric or asymmetric isotopomer, and the ro-vibrational symmetry. These data complement the data for the singly substituted ozone isotopomers (*i.e.* $^{16}O^{18}O^{16}O$ and $^{16}O^{16}O^{18}O$) reported in a recent work.[35] In what follows, the data for both singly and doubly substituted ozone molecules are fitted and analyzed together, as one comprehensive set of results.

## C. Fitting and analysis of the ro-vibrational spectra

In order to compare and contrast the spectra of symmetric and asymmetric ozone molecules we fitted their rotational energy levels using the following expression:

$$E_{\text{rot}}(J, \Lambda, p) = E_{\text{vib}} + \frac{A+B}{2} J(J+1) + \left(C - \frac{A+B}{2}\right)\Lambda^2 + (-1)^{J+\Lambda+p} \frac{\Delta W(J, \Lambda)}{2}. \quad (43)$$

The first term corresponds to the vibrational energy, the next two terms add rotational energy of the symmetric top rotor (parity-independent), and the last term is responsible for the splitting between the two parities, where the absolute value of the splitting is given by Wang's formula through binomial coefficients:[40,41]

$$\Delta W(J, \Lambda) = 8(C - A) \binom{J+\Lambda}{\Lambda}\binom{J}{\Lambda}\Lambda^2 \left(\frac{\beta}{8}\right)^\Lambda (1-\beta)^{-1}, \quad (44)$$

where

$$\beta = \frac{A - B}{2C - A - B} \quad (45)$$

is used to characterize the degree of asymmetry of the rotor.

First, we tried to fit the rotational spectrum of the ground vibrational state (0,0,0) in each ozone isotopomer by the symmetric-top rotor formula, with the parity splitting neglected, *i.e.* by setting $\Delta W = 0$ in Eq. (43). The results of such fitting are presented in Table 1. The first row shows the fitted values of the vibrational energy for the ground state of each molecule. The values in parenthesis are given for comparison and correspond to the exact vibrational energies, computed in this work (see *Supplemental Information*). The next two rows report the values of the fitting coefficients $(A + B)/2$ and $C$ for different isotopomers. Experimental data[42] are given in parenthesis for comparison. In all cases the fitted values of $(A + B)/2$ are in perfect agreement with the experimental data, while the fitted values of $C$ indicate differences on the order of 0.01 cm$^{-1}$. The last row gives the residual mean square error (RMSE) for a given fit, and those values are on the order of 0.1 cm$^{-1}$ for these fits.

Next, we fitted the rotational spectrum of the ground vibrational state in each ozone isotopomer using the fully relaxed version of Eq. (43). The values of fitting coefficients for this case are given in Table 2, which is structured in the same way as Table 1. In contrast to the Table 1, these fits correspond to the asymmetric-top rotor molecules and allow one to determine the values of $A$ and $B$ separately, based on the magnitudes of parity splittings. The non-zero difference between $A$ and $B$ also permits us to determine the value of asymmetry parameter $\beta$ for each isotopomer of ozone, reported in the fifth row of the Table 2. Note that when the $\Delta W$ parameter in Eq. (43) is relaxed, the values of RMSE are reduced by an order of magnitude, to about 0.01 cm$^{-1}$, which means that the quality of the fit of the data is significantly improved.

In different isotopomers and isotopologues of ozone the values of rotational constants are similar, roughly equal to $A \approx 0.42$ cm$^{-1}$, $B \approx 0.37$ cm$^{-1}$ and $C \approx 3.3$ cm$^{-1}$, with differences on the order of ±5% due to isotopic substitutions (see Table 2). These numbers satisfy reasonably well the condition of the symmetric top rotor approximation, $A \approx B \ll C$, which was frequently used in the past to ease calculations but is avoided here, in order to reach the new higher level of accuracy. The fitted values of $A$ and $B$ match the experimental values precisely for all molecules, with the exception of 0.001 cm$^{-1}$ difference for $B$ in the case of $^{18}O^{18}O^{16}O$. The fitted values of $C$ deviate from the corresponding experimental measurements only by 0.01 cm$^{-1}$ (see Table 2). The ground ro-vibrational energies predicted by these fits are also in good agreement with the results of the exact calculations (reported in *Supplementary Information*), all higher by only about 0.007

cm$^{-1}$. This excellent agreement with experimental results serves as another benchmark test for the accuracy of the code developed here.

The values of the asymmetry parameter approach $\beta \approx 0.01$ for all isotopomers. One can see that in the case of single isotopic substitution the symmetric ozone molecule $^{16}$O$^{18}$O$^{16}$O has slightly higher value of $\beta$ than the asymmetric molecule $^{16}$O$^{16}$O$^{18}$O. But, in case of the double substitution the behavior is reversed: now the asymmetric molecule $^{18}$O$^{18}$O$^{16}$O demonstrates slightly higher values of $\beta$, compared to the symmetric molecule $^{18}$O$^{16}$O$^{18}$O (see Table 2).

These trends are further explored in Figure 3, where we collected the values of parity splittings for the cases of $\Lambda = 1$ and $\Lambda = 2$, for each isotopomer of ozone considered here. Roughly, for $\Lambda = 1$ the splittings are on the order of $\Delta W \approx 0.04$ cm$^{-1}$ for $J = 1$, and they are increased tenfold when the rotational excitation is raised to $J = 4$, reaching $\Delta W \approx 0.4$ cm$^{-1}$. In the case of $\Lambda = 2$, the splittings are about two orders of magnitude smaller, starting from $\Delta W \approx 0.0005$ cm$^{-1}$ for $J = 2$ and reaching about $\Delta W \approx 0.02$ cm$^{-1}$ for $J = 5$. (The data presented in Figure 3 are also reported in Tables S1 and S2 of *Supplementary Information*.) From this figure one can clearly see that symmetric and asymmetric ozone molecules behave differently in the cases of singly and doubly substituted ozone. Namely, in the case of single substitution the splitting is larger for the symmetric isotopomer, while in the case of double substitution the splitting is larger for the asymmetric isotopomer.

In order to illuminate the effect of vibrational excitation, we modified Eq. (43) by expressing $E_{\text{vib}}$ through the second order Dunham expansion:

$$E_{\text{vib}}(v_1, v_2, v_3) = E_{\text{elec}} + \sum_{i=1}^{3} \omega_i \left( v_i + \frac{1}{2} \right) + \sum_{i,j=1}^{3} \delta_{ij} \left( v_i + \frac{1}{2} \right) \left( v_j + \frac{1}{2} \right). \tag{46}$$

The first term of Eq. (46) is the lowest energy on the PES, the bottom of the well. The next term adds harmonic contribution from each mode (3 normal modes total in case of ozone) and the last term adds the intra-mode and inter-mode anharmonicities.

First, we used Eqs. (43)-(46) to fit only the ro-vibrational states with no more than one quantum of the vibrational excitation in each mode, assuming a harmonic oscillator model, *i.e.* setting all $\delta_{ij} = 0$. The results of such fitting are presented in Table 3 for all isotopomers of ozone considered here. Now the first row represents electronic energy $E_{\text{elec}}$ relative to the dissociation limit. For comparison, the energy values at the minimum energy point on the PES are given in parenthesis (different in the singly and doubly substituted ozone molecules, since the dissociation energy includes zero-point energy of the heaviest diatomic fragment, which is $^{16}O^{18}O$ in the case of the singly substituted ozone but is $^{18}O^{18}O$ in the case of the doubly substituted ozone). The next three rows of Table 3 report the fitted values of harmonic frequencies $\omega_1$, $\omega_2$ and $\omega_3$. For comparison, experimental values of the fundamental excitation energies[43] are given in parenthesis for each molecule. These data demonstrate a very good agreement between theory and experiment, with differences of only ~ 6 cm$^{-1}$ in all modes of all isotopomers. The next three rows of Table 3 list the rotational constants $A$, $B$ and $C$ derived from this ro-vibrational fit. Their values are similar to the ones given in Tables 1 and 2, but not exactly the same, which indicates that vibrational excitation has some effect on the rotational spectrum.

To explore this question in detail, we carried out the fits of rotational spectra using Eqs. (43)-(45) separately for the first excited vibrational state of each mode: (001), (010) and (100). The resultant fitting parameters are collected in the Tables S3, S4 and S5 of the *Supplemental Information*. These data indicate that excitation of the bending vibration mode, does not affect asymmetry of the rotor. Namely, the values of asymmetry parameter $\beta$ for the first excited vibrational state (0,1,0) of all isotopomers appear to be very similar to those of the ground state (0,0,0) reported in the Table 2. However, we found that the excitations of symmetric and asymmetric stretching modes affect the asymmetry of the rotor. Interestingly, we found that excitation of an asymmetric stretch state (0,0,1) increases the values of parameter $\beta$ for all ozone isotopomers (making the rotor more asymmetric) while the excitation of a symmetric stretch state (1,0,0) decreases the values of parameter $\beta$ for all ozone isotopomers (making the rotor less asymmetric). These effects are not negligible, on the order of 10% of the $\beta$ values.

Overall, the accuracy of the common ro-vibrational fit, Eqs. (43)-(46), is lower compared to the rotational fits of individual vibrational states, but this fit is still reliable. In the last row of Table 3 we listed RMSE for different ozone isotopomers, and those numbers are around 0.22 cm$^{-1}$.

Finally, we used Eq. (46) without restrictions on anharmonicities to fit the rovibrational states with no more than 2 quanta of excitation, cumulatively across all modes, which includes overtones and combination bands (10 vibrational states total). For these fits the values of RMSE increase again but not critically, reaching 0.35 cm$^{-1}$ on average for all isotopic substitutions considered here. This number is not large, considering the span of the fitted spectrum of roughly 2000 cm$^{-1}$ which fills about a quarter of the potential energy well in ozone on its way to the dissociation towards O + O$_2$.

The first row in Table 4 shows excellent agreement between the fitted and the actual electronic energies, with the average deviation of about 2 cm$^{-1}$. The values of harmonic frequencies $\omega_1$, $\omega_2$ and $\omega_3$ in Table 4 should not be mixed with excitation energies, and should not be directly compared to the experimental data given in Table 3 since those numbers do not take into account anharmonicity effects. Analysis of the intra-mode anharmonicity parameters in Table 4 indicates that the bending mode is the least anharmonic of all, with $\delta_{22} \approx -1.2$ cm$^{-1}$, while the asymmetric-stretching mode is the most anharmonic of all, with more than ten times larger anharmonicity parameter of about $\delta_{33} \approx -14$ cm$^{-1}$. Both of these characteristics change little across the four isotopic substitutions considered here, indicating similar values for symmetric and asymmetric ozone molecules with single and double isotopic substitutions. But we found that the symmetric-stretching mode in ozone has its own interesting property: This mode is less anharmonic in symmetric ozone molecules with $\delta_{11} \approx -2.7$ cm$^{-1}$ and is more anharmonic in asymmetric ozone molecules with $\delta_{11} \approx -4.9$ cm$^{-1}$, and this large difference is systematically present in both singly and doubly-substituted ozone species. The inter-mode anharmonicity parameters $\delta_{12} \approx -7.5$ cm$^{-1}$ and $\delta_{23} \approx -15$ cm$^{-1}$ remain roughly the same across the four isotopic substitutions considered here. But, the value of $\delta_{13}$ behaves differently: It is larger in symmetric ozone molecules, $\delta_{13} \approx -32$ cm$^{-1}$ and is smaller in asymmetric ozone molecules, $\delta_{13} \approx -25$ cm$^{-1}$, and this appreciable difference is systematically present in both singly and doubly-substituted ozone species.

### D. Ro-vibrational partition functions

Excellent agreement of the fitted spectroscopic constants with the experimental results, together with the low values of RMSE of the fits in Tables 1-4, permit us to use Eqs. (43-46) to estimate the behavior of the spectrum of ozone molecules at larger values of $J$ that have not yet been calculated explicitly. Figure 4 shows extrapolation of the parity splittings for the ground vibrational state of the singly substituted isotopomers ($^{16}$O$^{18}$O$^{16}$O and $^{16}$O$^{16}$O$^{18}$O) as a function of $J$ for different values of $\Lambda$. The fitted data points, available in the range $1 \leq J \leq 5$, are shown by symbols. Solid and dashed lines correspond to the analytic fits of these data for symmetric $^{16}$O$^{18}$O$^{16}$O and asymmetric $^{16}$O$^{16}$O$^{18}$O isotopomers, respectively. The fits are extended to extrapolate up to $J = 50$. The curves corresponding to $1 \leq \Lambda \leq 5$ are labelled explicitly in the picture; the curves for $\Lambda > 5$ can be easily identified using the overall trend. Figure 5 shows similar data for the doubly substituted isotopomers, symmetric $^{18}O^{16}O^{18}O$ and asymmetric $^{18}O^{18}O^{16}O$.

From the Figures 4 and 5, and from Eq. (44), one can see that for the low values of $J$ the splittings between different parities decrease exponentially as a function of $\Lambda$ but they increase as a function of $J$, as $O(J^{2\Lambda})$. Thus, the curves corresponding to the higher values of $\Lambda$ start lower in Figures 4 and 5, but grow faster and eventually cross the curves corresponding to the lower values of $\Lambda$. This is indeed what we can see at $J \approx 30$, where $\Lambda = 1$ crosses the $\Lambda = 2$ curve, and at $J \approx 50$, where the $\Lambda = 2$ curve is crossed by $\Lambda = 3$. The analytical fits allow us to predict that in the region of $J = 50$ the states with $\Lambda = 1$ to 5 are expected to have splittings above 1 cm$^{-1}$.

As for the symmetric vs asymmetric molecule behavior, the trends reported in Figure 3 for the low values of $J$ are expected to hold for higher values of $J$ as well. Namely, Figures 4 and 5 indicate that the splittings of $^{16}O^{18}O^{16}O$ are greater than those of $^{16}O^{16}O^{18}O$ in the whole range of the considered values of $J$, while for the doubly substituted isotopomers the behavior is just the opposite, i.e. the splittings for $^{18}O^{18}O^{16}O$ are greater than those of $^{18}O^{16}O^{18}O$. This order is not expected to change for any value of $J$ and $\Lambda$ due to the way the splittings depend on them in Eq. (44), although the absolute value of difference between the splittings in the symmetric and asymmetric molecules grows as a function of $J$, which can be clearly seen in the case of $\Lambda = 5$ in Figures 4 and 5.

Extrapolation of the spectra toward large values of $J$ can also be used to compute the ro-vibrational partition functions $Q_{\text{asym}}$ and $Q_{\text{sym}}$ for asymmetric and symmetric ozone molecules considered in this work. These, in turn, can be used to determine the ratio of the number of states in asymmetric and symmetric ozone molecules, $R = Q_{\text{asym}}/Q_{\text{sym}}$, which may deviate from the statistical value of $R = 2$. Figure 6 summarizes our results for the singly and doubly substituted ozone in the range of temperatures relevant to the stratosphere and the laboratory studies. The solid blue and red lines give the values of $R = Q_{\text{asym}}/Q_{\text{sym}}$ for the singly and doubly substituted ozone molecules respectively (calculated from their extrapolated spectra). In each case the spectrum was fitted with Eqs. (43)-(46), using the rovibrational states with $0 \leq J \leq 5$ and up to 2 quanta of vibrational excitation, and extrapolated up to the energy ~4000 cm$^{-1}$ above the bottom of the potential energy well. One can see that in the singly substituted ozone molecule the ratio of the partition functions deviates from the statistical value of $R = 2$ by about +0.05 (which is on the order of 2.5%) in the whole range of the considered temperatures, while in the doubly substituted case the same deviation occurs in the opposite direction, –0.05. Interestingly, the singly-substituted and the doubly-substituted ozone molecules behave differently, and the difference of $R$ values for them is on the order of 0.1, which is a substantial deviation from the statistical value of $R = 2$.

The dashed red and blue lines in Figure 6 are given to demonstrate the effect of parity splittings on the value of the ratio $R = Q_{\text{asym}}/Q_{\text{sym}}$. These dashed lines were obtained using the fits of the spectra by a simplified expression, with fixed $\Delta W = 0$ in Eq. (43), which corresponds to the symmetric-top rotor approximation with splittings neglected. One can see that at low temperatures the effect of parity splittings is negligible, since only the low levels of rotational excitations are assessible, where the values of parity splittings remain small. For higher temperatures the effect of splittings on $R = Q_{\text{asym}}/Q_{\text{sym}}$ becomes visible in Figure 6, but is still relatively small (on the order of 0.005), an order of magnitude smaller than the effect of the single vs double isotopic substitutions, as emphasized by Figure 6.

Gray lines in the background of Figure 6 were obtained using the rotational partition functions $Q_{\text{asym}}$ and $Q_{\text{sym}}$ of the ground vibrational state (0,0,0), without including any excited vibrational states. These are given to illustrate the effect of vibrational excitation. As before, the dashed gray lines correspond to the symmetric-top rotor case, when the parity splittings are neglected. We can see that inclusion of the vibrational excitations leads to some shape of the $R(T)$ dependencies and becomes more important at higher temperatures. Without vibrations, the values of the ratios $R = Q_{\text{asym}}/Q_{\text{sym}}$ remain nearly constant through the considered temperature range.

## IV. CONCLUSIONS

In this paper we developed theory and tested a new code for the efficient calculation of coupled rotational-vibrational states in triatomic molecules using hyper-spherical coordinates and taking into account all terms of the Hamiltonian operator, including the asymmetric-top rotor couplings and the Coriolis couplings. Concise final formulae were derived for the efficient calculations of matrix elements, for construction of the Hamiltonian matrix, for expressing the total ro-vibrational wavefunction, for the assignment of quantum numbers to the computed eigenstates, and finally for the identification of possible isotopomers of the molecule on the global PES (*e.g.,* symmetric *vs* asymmetric ozone). Our numerical approach is distinct from other available methods, since it uses an efficient combination of the FBR and DVR methods (taking advantage of adaptive grids which adjust to the shape of the PES) and reduces significantly the size of the Hamiltonian matrix (by constructing and truncating the locally-optimal basis sets at all levels of the calculations). Matrix-diagonalization is carried out using PARPACK,[44] whereas the rest of the code is parallelized using MPI, which makes it suitable for massively parallel execution on a variety of high-performance computing platforms.

First, this new code was rigorously tested by running a set of benchmark calculations using one of the existing well-tested computer codes (the APH3D program of Kendrick) and comparing the results of the two codes. Excellent agreement was found. Next, the new code was used to compute the coupled rotational-vibrational states of ozone molecule with double substitutions of $^{18}O$ isotope, which includes symmetric $^{18}O^{16}O^{18}O$ and asymmetric $^{18}O^{18}O^{16}O$ isotopomers. Here we covered the values of total angular momentum up to $J = 5$ with both inversion parities and all the rotational blocks $\Lambda \leq J$ (without any approximations). The range of covered vibrational energies extends up to ~4000 cm$^{-1}$ above the ground vibrational state, which includes up to five quanta of vibrational excitations and covers about one-half of the potential energy well in ozone. To the best of our knowledge such calculations have never been reported before for the rotationally excited doubly substituted ozone molecule.

Together with similar data recently reported for symmetric $^{16}O^{18}O^{16}O$ and asymmetric $^{16}O^{16}O^{18}O$ isotopomers of the singly substituted isotopologues, our new data enable a systematic analysis of isotope effects in the rotational-vibrational spectra of ozone. Namely, we checked whether it is reasonable to expect that, due to the Coriolis coupling effect, the asymmetric ozone isotopomers (singly substituted $^{16}O^{16}O^{18}O$ and doubly substituted $^{18}O^{18}O^{16}O$) would behave similar to each other but different from the symmetric ozone isotopomers (singly substituted $^{16}O^{18}O^{16}O$ and doubly substituted $^{18}O^{16}O^{18}O$), which in turn would also behave similar to each other. So far, we found no justification for this hypothesis. We found that for ozone the deviations of rotational constants from the standard symmetric-top-rotor behavior is affected by isotopic composition as much as it is affected by the symmetry of the molecule. For example, in the case of single isotopic substitution the value of the rotational asymmetry parameter $\beta$ appears to be smaller in asymmetric $^{16}O^{16}O^{18}O$ than it is in symmetric $^{16}O^{18}O^{16}O$, but, it is just opposite in the case of double substitution, where the value of the rotational asymmetry parameter $\beta$ is found to be larger in asymmetric $^{18}O^{18}O^{16}O$ than it is in symmetric $^{18}O^{16}O^{18}O$.

Another relevant feature, that has never been discussed in the literature on ozone before, is the value of parity splitting ($\Lambda$-doubling) due to the Coriolis coupling effect. These splittings, accurately captured by our calculations, were determined and examined here for $1 \leq \Lambda \leq 5$, for all four ozone isotopomers. We found that these splittings are affected by isotopic substitutions as much as they are affected by molecular symmetry, namely: in the case of single isotopic substitution the splittings are larger in symmetric ozone $^{16}O^{18}O^{16}O$, but in the case of double isotopic substitution the splittings are larger in asymmetric ozone $^{18}O^{18}O^{16}O$. Again, one can't claim that symmetry is a determining factor.

Then we checked how a "bulk" energy-averaged characteristic of the molecule, such as its rotational-vibrational partition function, is affected by the Coriolis coupling effect, and how much these partition functions are different in different isotopomers of ozone. Since it is expected that the number of allowed ro-vibrational states in asymmetric molecules would be twice larger than it is in symmetric molecules, we have chosen to use the ratio of partition functions for asymmetric and symmetric ozone molecules to serve as a useful metric: $R = Q_{\text{asym}}/Q_{\text{sym}}$. Its value is expected to be close to $R = 2$, so, any deviation would be considered as an isotope effect. We found, first of all, that for the temperatures below 500 K the effect of parity splittings on the ratio $R$ is very small and thus the role of the Coriolis coupling is negligible. We also found that the accurately computed value of this metric deviates from the expected statistical $R = 2$, but the direction of this deviation depends on the number of isotopic substitutions. Namely, in the singly substituted case the ratio $^{16}O^{16}O^{18}O/^{16}O^{18}O^{16}O$ is larger than expected, while in the doubly substituted case the ratio $^{18}O^{18}O^{16}O/^{18}O^{16}O^{18}O$ is smaller than expected, in both cases by approximately the same amount, ±0.05. Although by itself this is an interesting isotope-related phenomenon, this effect is relatively small, and is driven by masses, not by the symmetry.

Here we computed the rotational levels of ozone molecules only up to $J = 5$, fitted them with analytic expressions, and used those to estimate the rotational spectra for $J > 5$ (including the magnitudes of splittings due to the Coriolis coupling effect). This extrapolation scheme is expected to be reasonably accurate, but it would certainly be better to have accurate results for higher values of $J$. These calculations are progressively ongoing. They take a significant amount of computer and wall clock time, and thus will be reported elsewhere. The code developed in this work is general and can be used for predictions of the rotational-vibrational states of any triatomic molecule. The authors plan to make this code publicly available to the community in the near future.

**SUPPLEMENTARY MATERIAL**

Section A of the *Supplementary Information* includes tables with the data for Figure 3.

Section B of the *Supplementary Information* describes an alternative fitting approach, where each vibrational state is allowed to have its own set of rotational constants. Several examples of such fits are also given.

Section C of the *Supplementary Information* provides details about the values of constants used in this work and includes all ro-vibrational states calculated in this work with their respective assignments.

**ACKNOWLEDGMENTS**

This research was supported by the NSF AGS program Grant No. AGS-1920523. We used resources of the National Energy Research Scientific Computing Center, which is supported by the Office of Science of the U.S. Department of Energy under Contract No. DE-AC02-5CH11231. IG acknowledges the support of Schmitt Fellowship at Marquette. AT and BKK acknowledge that part of this work was done under the auspices of the US Department of Energy under Project No. 20180066DR of the Laboratory Directed Research and Development Program at Los Alamos National Laboratory. Los Alamos National Laboratory is operated by Triad National Security, LLC, for the National Nuclear Security Administration of the U.S. Department of Energy (Contract No. 89233218CNA000001).

**TABLES**

**Table 1.** Least squares fitting coefficients (in cm$^{-1}$) of Eq. (43), where the parity splitting term $\Delta W$ is set to 0, computed using all rotational states with $0 \leq J \leq 5$ of the ground vibrational state for various ozone isotopomers. The numbers in parenthesis are experimental spectroscopic constants,[42] or, in case of $E_{\text{vib}}$, the accurately computed energies of the ground vibrational state. Energy is defined with respect to the lower dissociation channel of the corresponding isotopomer.

| Parameter | $^{16}O^{18}O^{16}O$ | $^{16}O^{16}O^{18}O$ | $^{18}O^{16}O^{18}O$ | $^{18}O^{18}O^{16}O$ |
|---|---|---|---|---|
| $E_{\text{vib}}$ | -8629.717 | -8615.466 | -8617.638 | -8632.133 |
|  | (-8629.724) | (-8615.474) | (-8617.645) | (-8632.140) |
| $(A+B)/2$ | 0.418 | 0.397 | 0.375 | 0.395 |
|  | (0.418) | (0.397) | (0.375) | (0.396) |
| $C$ | 3.300 | 3.498 | 3.432 | 3.235 |
|  | (3.290) | (3.488) | (3.422) | (3.225) |
| RMSE | 0.123 | 0.105 | 0.0954 | 0.112 |

**Table 2.** Least squares fitting coefficients (in cm$^{-1}$) of Eq. (43), computed using all rotational states with $0 \leq J \leq 5$ of the ground vibrational state for various ozone isotopomers. The numbers in parenthesis are experimental spectroscopic constants,[42] or, in case of $E_{\text{vib}}$, the accurately computed energy of the ground vibrational state. Energy is defined with respect to the lower dissociation channel of the corresponding isotopomer.

| Parameter | $^{16}O^{18}O^{16}O$ | $^{16}O^{16}O^{18}O$ | $^{18}O^{16}O^{18}O$ | $^{18}O^{18}O^{16}O$ |
|---|---|---|---|---|
| $E_{\text{vib}}$ | -8629.717 | -8615.466 | -8617.638 | -8632.133 |
|  | (-8629.724) | (-8615.474) | (-8617.645) | (-8632.140) |
| $A$ | 0.445 | 0.420 | 0.396 | 0.420 |
|  | (0.445) | (0.420) | (0.396) | (0.420) |
| $B$ | 0.391 | 0.374 | 0.354 | 0.371 |
|  | (0.391) | (0.374) | (0.354) | (0.372) |
| $C$ | 3.300 | 3.498 | 3.432 | 3.235 |
|  | (3.290) | (3.488) | (3.422) | (3.225) |
| $\beta$ | 9.36x10$^{-3}$ | 7.40x10$^{-3}$ | 6.82x10$^{-3}$ | 8.66x10$^{-3}$ |
|  | (9.40x10$^{-3}$) | (7.44x10$^{-3}$) | (6.87x10$^{-3}$) | (8.48x10$^{-3}$) |
| RMSE | 0.0128 | 0.0126 | 0.0118 | 0.0118 |

**Table 3.** Least squares fitting coefficients (in cm$^{-1}$) of Eq. (43) and (46), where the anharmonicity terms $\delta_{ij}$ are set to 0, computed using all rotational states with $0 \leq J \leq 5$ and vibrational states with up to 1 quanta of excitation (4 vibrational states total) for various ozone isotopomers. The numbers in parenthesis are experimental spectroscopic constants[42,43] or, in case of $E_{elec}$, the actual lowest energy of the PES. Energy is defined with respect to the lower dissociation channel of the corresponding isotopomer.

| Parameter | $^{16}O^{18}O^{16}O$ | $^{16}O^{16}O^{18}O$ | $^{18}O^{16}O^{18}O$ | $^{18}O^{18}O^{16}O$ |
|---|---|---|---|---|
| $E_{elec}$ | -10015 (-10044) | -10014 (-10044) | -9995 (-10024) | -9995 (-10024) |
| $\omega_1$ | 1068 (1074) | 1085 (1090) | 1066 (1072) | 1055 (1061) |
| $\omega_2$ | 687.7 (696.3) | 679.2 (684.6) | 662.8 (668.1) | 672.1 (677.5) |
| $\omega_3$ | 1015 (1008) | 1034 (1028) | 1025 (1019) | 999.3 (993.9) |
| $A$ | 0.443 (0.445) | 0.418 (0.420) | 0.394 (0.396) | 0.418 (0.420) |
| $B$ | 0.389 (0.391) | 0.372 (0.374) | 0.352 (0.354) | 0.369 (0.372) |
| $C$ | 3.301 (3.290) | 3.499 (3.488) | 3.432 (3.422) | 3.235 (3.225) |
| RMSE | 0.236 | 0.233 | 0.236 | 0.216 |

**Table 4.** Least squares fitting coefficients (in cm$^{-1}$) of Eq. (43) and (46), computed using all rotational states with $0 \leq J \leq 5$ and vibrational states with up to 2 quanta of excitation (cumulatively on all modes, 11 vibrational states total) for various ozone isotopomers. The numbers in parenthesis are experimental spectroscopic constants[42,43] or, in case of $E_{elec}$, the actual lowest energy of the PES. Energy is defined with respect to the lower dissociation channel of the corresponding isotopomer.

| Parameter | $^{16}O^{18}O^{16}O$ | $^{16}O^{16}O^{18}O$ | $^{18}O^{16}O^{18}O$ | $^{18}O^{18}O^{16}O$ |
|---|---|---|---|---|
| $E_{elec}$ | -10042 (-10044) | -10042 (-10044) | -10021 (-10024) | -10022 (-10024) |
| $\omega_1$ | 1094 | 1112 | 1090 | 1082 |
| $\omega_2$ | 701.5 | 693.3 | 676.4 | 685.4 |
| $\omega_3$ | 1064 | 1084 | 1076 | 1046 |
| $\delta_{11}$ | -2.919 | -4.861 | -2.484 | -4.865 |
| $\delta_{22}$ | -1.308 | -1.283 | -1.210 | -1.249 |
| $\delta_{33}$ | -12.83 | -14.95 | -13.82 | -13.70 |
| $\delta_{12}$ | -7.419 | -7.850 | -7.140 | -7.490 |
| $\delta_{13}$ | -33.00 | -26.26 | -31.30 | -25.35 |
| $\delta_{23}$ | -14.87 | -15.21 | -15.26 | -14.17 |
| $A$ | 0.441 (0.445) | 0.416 (0.420) | 0.392 (0.396) | 0.416 (0.420) |
| $B$ | 0.387 (0.391) | 0.370 (0.374) | 0.350 (0.354) | 0.367 (0.372) |
| $C$ | 3.302 (3.290) | 3.500 (3.488) | 3.433 (3.422) | 3.236 (3.225) |
| RMSE | 0.363 | 0.361 | 0.357 | 0.336 |

**FIGURES**

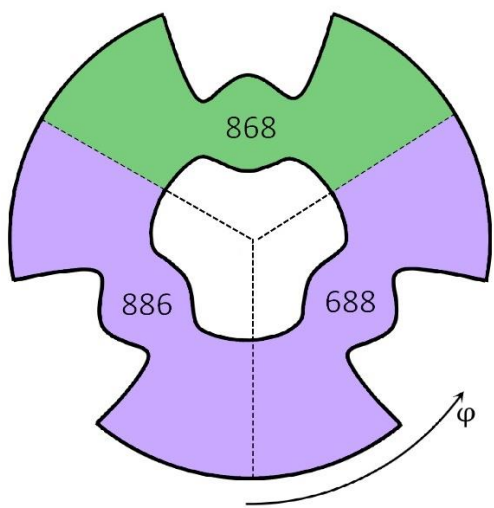

Figure 1. Schematic representation of the PES of ozone in the hyper-spherical coordinates, to illustrate differences between symmetric and asymmetric isotopomers. Three covalent wells are labelled as "886", "688" and "868", where "6" and "8" stand for $^{16}O$ and $^{18}O$ respectively. Pink and orange colors mark the regions of the PES conditionally associated with the symmetric and asymmetric ozone molecules, respectively.

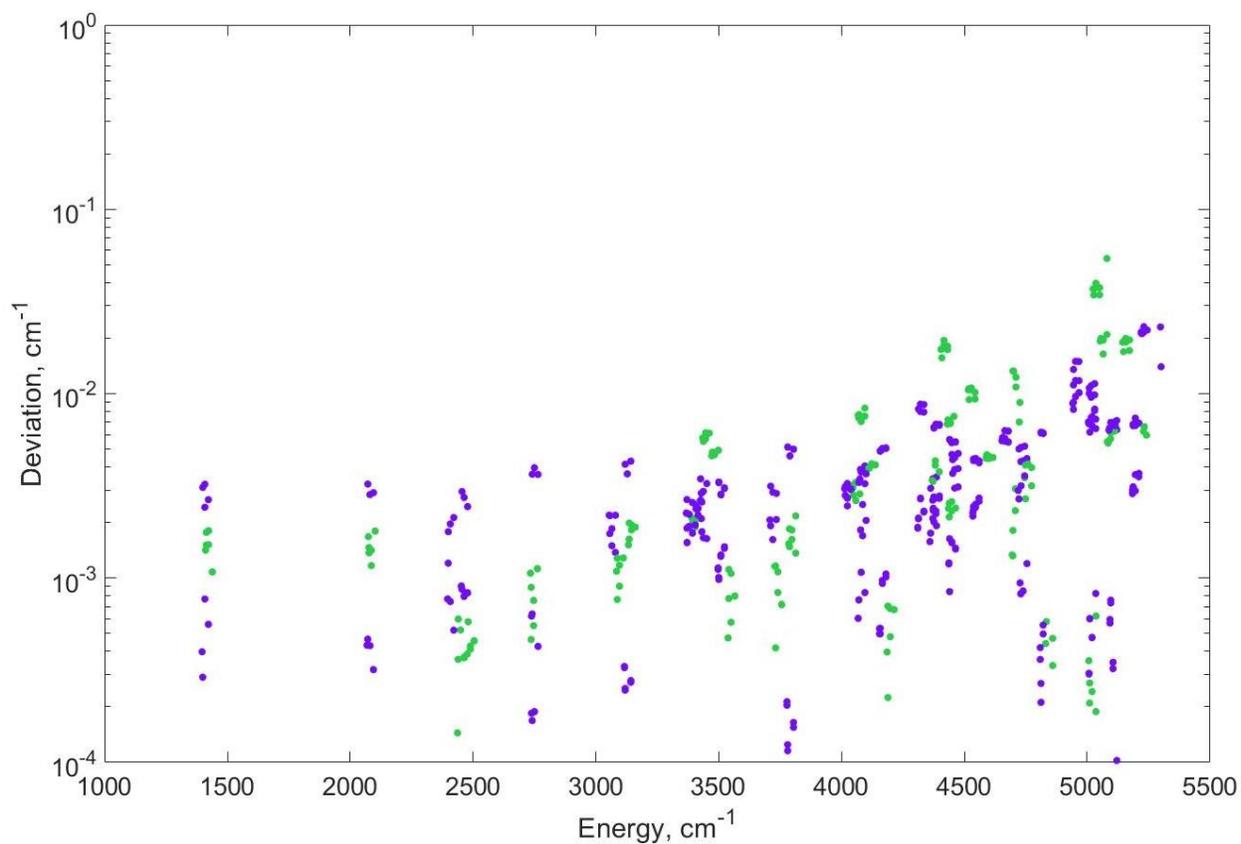

Figure 2. Absolute values of energy difference between the rovibrational states of ozone calculated using the code developed here and the code of Kendrick.[38] In this example the states with both values of inversion parity ($p = 0$ and $p = 1$) are shown for the total angular momentum $J = 3$ of doubly substituted ozone molecule. The states of both $^{18}O^{16}O^{18}O$ (green) and $^{18}O^{18}O^{16}O$ (violet) are included. Horizontal axis gives energy relative to the bottom of the well.

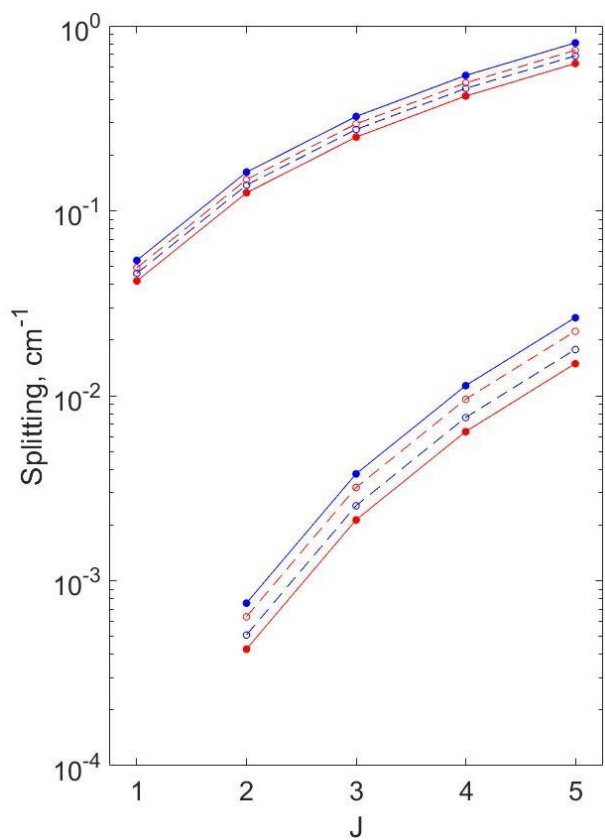

Figure 3. Absolute values of parity splittings in ozone for $\Lambda = 1$ and $\Lambda = 2$, as a function of $J$. Filled and empty symbols correspond to the values of splittings computed directly from the ro-vibrational energies (reported in *Supplementary Information*) for symmetric and asymmetric isotopomers, respectively. Solid and dashed lines show the analytic fit of these data by Eq. (44) for symmetric and asymmetric isotopomers, respectively. The blue and red colors correspond to singly and doubly substituted ozone molecules.

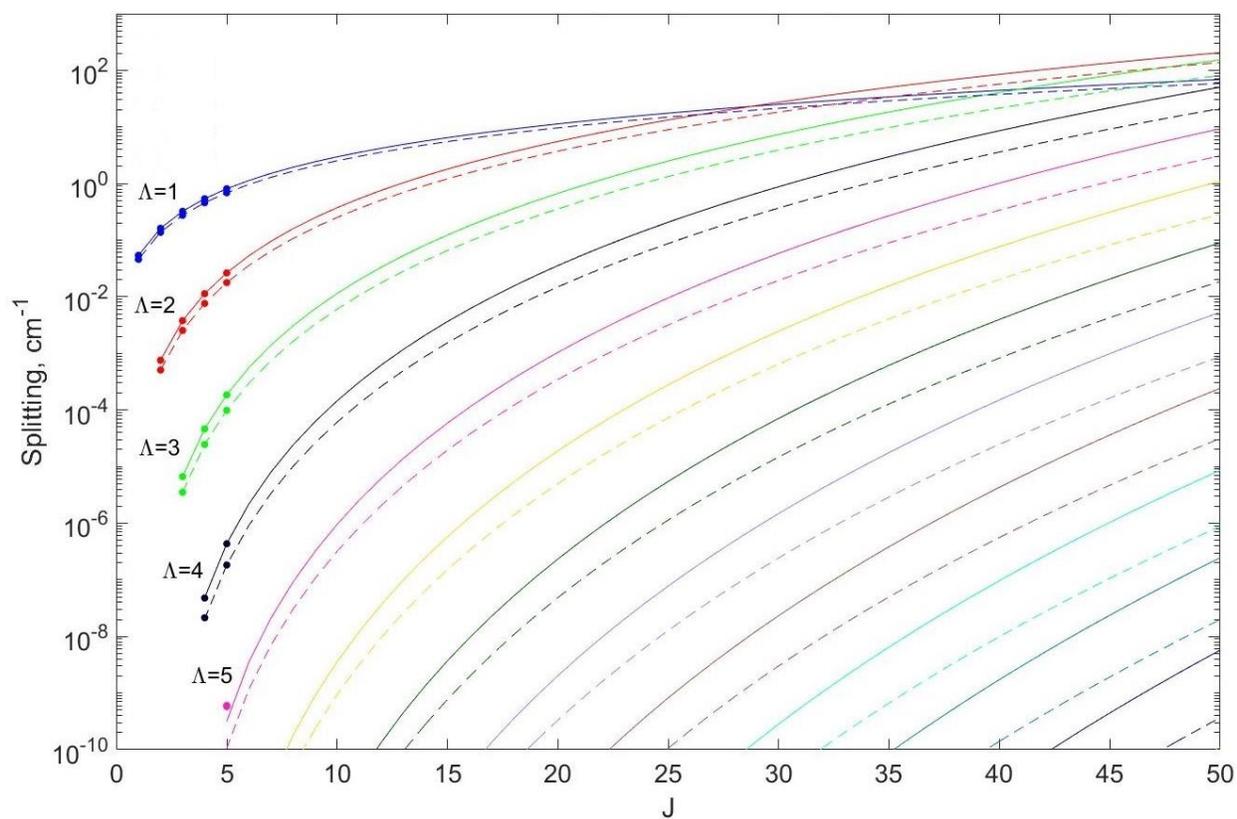

Figure 4. Extrapolation of parity splittings for $^{16}O^{18}O^{16}O$ (solid line) and $^{16}O^{16}O^{18}O$ (dashed line) as a function of $J$. Symbols mark exact values of splittings calculated in this work. Different values of $\Lambda$ are shown by different colors.

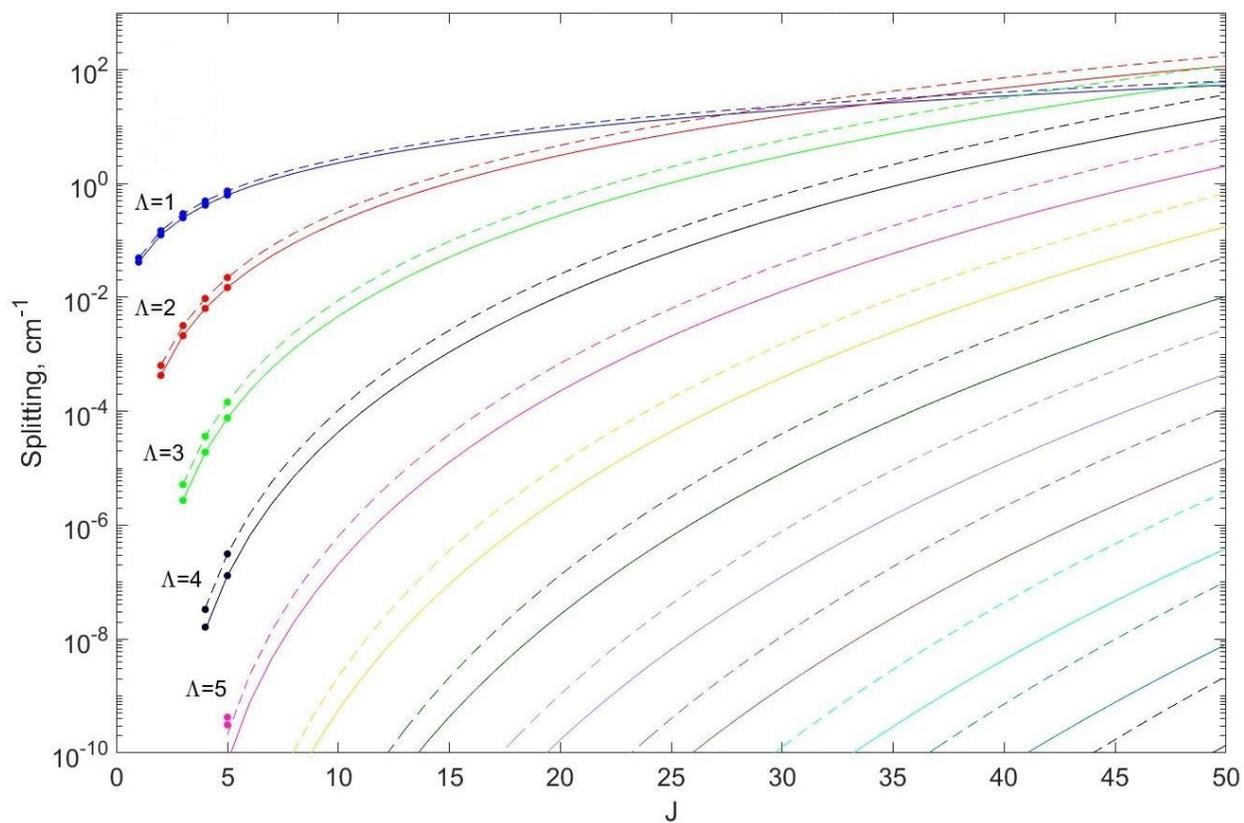

Figure 5. Extrapolation of parity splittings for $^{18}O^{16}O^{18}O$ (solid line) and $^{18}O^{18}O^{16}O$ (dashed line) as a function of $J$. Symbols mark exact values of splittings calculated in this work. Different values of $\Lambda$ are shown by different colors.

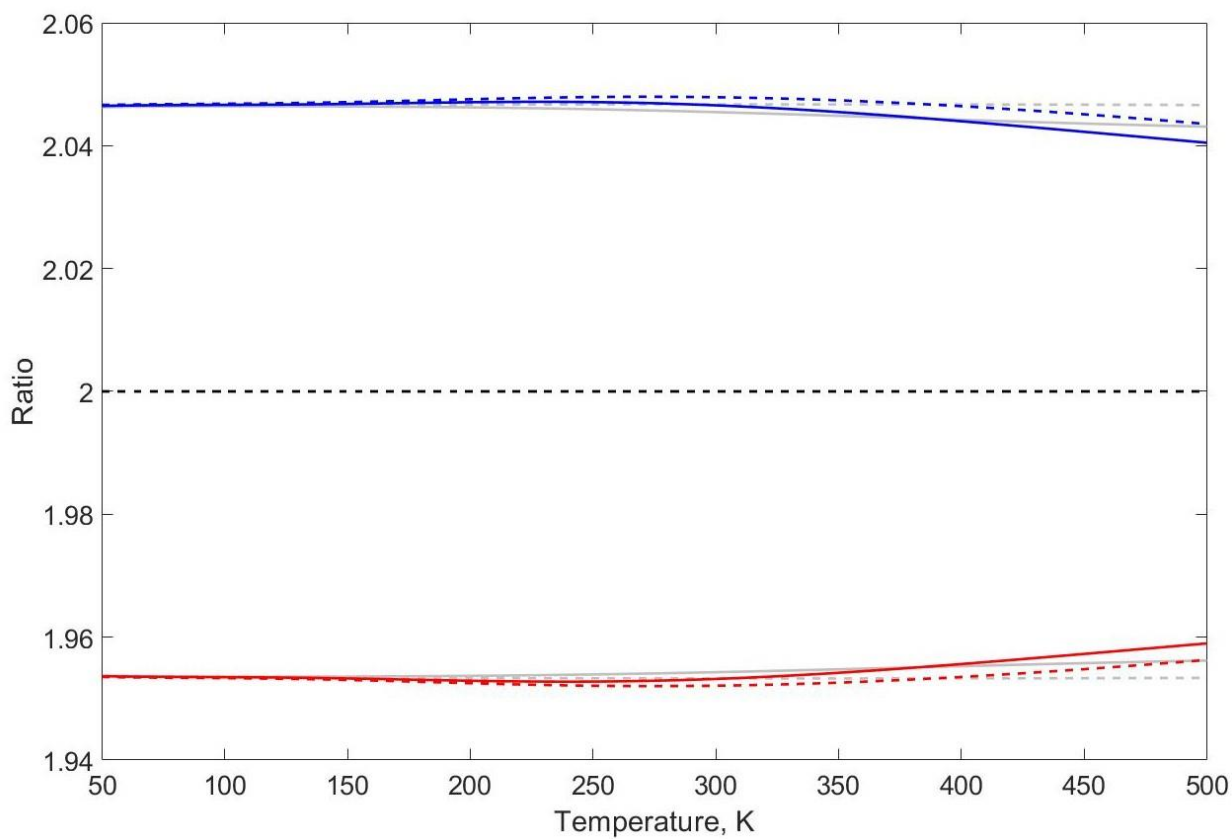

Figure 6. The ratio of partition functions of asymmetric and symmetric isotopomers of ozone. The solid blue (red) color corresponds to the singly (doubly) substituted isotopologues of ozone. The dashed lines correspond to a symmetric top rotor approximation, when the parity splittings are neglected. The gray lines in the background show analogous result calculated based on purely rotational spectrum, without inclusion of any vibrationally excited states.

## APPENDIX A. VIBRATIONAL BASIS SETS

The DVR basis functions for the hyper-radius $\rho$ and hyper-angle $\theta$ are:

$$h_n(\rho_i) = \begin{cases} \frac{1}{\sqrt{\Delta\rho}}, \rho_i = \rho_n, \\ 0, \rho_i \neq \rho_n. \end{cases} \quad (A1)$$

$$g_l(\theta_i) = \begin{cases} \frac{1}{\sqrt{\Delta\theta}}, \theta_i = \theta_l, \\ 0, \theta_i \neq \theta_l. \end{cases} \quad (A2)$$

where $\Delta\rho$ and $\Delta\theta$ are the step sizes on grids along $\rho$ and $\theta$ respectively. In case of $\theta$, the points $\theta_i$ are simply equidistant with a fixed step size. In the case of $\rho$, the placement of points is optimized based on the shape of the PES in a way that puts more points in the region of deep covalent well and fewer points in the shallow van der Waals interaction (asymptotic) region. This reduces the number of points necessary for the targeted accuracy. Even though the spacing between points is not equidistant, one can still work with it as if it was equidistant by using a mapping procedure. The details of this can be found elsewhere.[39]

Since the points along $\theta$ are equidistant, one can use an analytical expression to evaluate the kinetic energy matrix elements as follows:[39]

$$\langle g_l | \hat{T}_\theta^n | g_{l'} \rangle = \begin{cases} \dfrac{\pi^2}{(\theta_{max} - \theta_{min})^2} \dfrac{L^2 + 2}{6} & \text{if } l = l', \\ (-1)^{l-l'} \dfrac{\pi^2}{(\theta_{max} - \theta_{min})^2} \dfrac{1}{\sin^2\left(\dfrac{(l-l')\pi}{L}\right)} & \text{if } l \neq l'. \end{cases} \quad (A3)$$

For $\varphi$, we use a set of normalized cosine (vibrational symmetry A$_1$, labelled by "+" for symmetric) or sine (vibrational symmetry B$_1$, labelled by "−" for antisymmetric) functions:

$$f_m^+(\varphi) = \frac{1}{\sqrt{\pi(\delta_{m0} + 1)}} \cos(m\varphi), m = 0 \ldots M - 1, \quad (A4)$$

$$f_m^-(\varphi) = \frac{1}{\sqrt{\pi}} \sin(m\varphi) \ m = 1 \ldots M. \quad (A5)$$

If symmetry does not matter within a given context, we simply omit the symmetry label and write $f_m(\varphi)$ to eliminate ± subscript for clarity of notation.

The values of $\langle f_m | \frac{d}{d\varphi} | f_{m'} \rangle$ depend on the mutual symmetry of two functions and two cases are possible:

$$\langle f_m^\pm | \frac{d}{d\varphi} | f_{m'}^\pm \rangle = \mp m \langle f_m^\pm | f_{m'}^\mp \rangle = 0, \tag{A6}$$

$$\langle f_m^\pm | \frac{d}{d\varphi} | f_{m'}^\mp \rangle = \pm m \langle f_m^\pm | f_{m'}^\pm \rangle = \pm m \delta_{mm'}. \tag{A7}$$

As one can see, the integral is non-zero only when the functions of different symmetries are supplied. The absence of the Coriolis coupling between the functions $f_m$ of the same symmetry (together with the other features of the Hamiltonian matrix structure) makes it possible to separate the Hamiltonian matrix into 2 submatrices and diagonalize them separately, as shown in Figure A1.

|        | $A_1^0$ | $B_1^0$ | $A_1^1$ | $B_1^1$ | $A_1^2$ | $B_1^2$ | $A_1^3$ | $B_1^3$ |
|--------|---------|---------|---------|---------|---------|---------|---------|---------|
| $A_1^0$ | S       |         |         |         | C       | A       |         |         |
| $B_1^0$ |         | S       | C       |         |         | A       |         |         |
| $A_1^1$ |         | C       | SA      |         |         |         | C       | A       |
| $B_1^1$ | C       |         |         | SA      | C       |         |         | A       |
| $A_1^2$ | A       |         |         | C       | S       |         |         | C       |
| $B_1^2$ |         | A       | C       |         |         | S       | C       |         |
| $A_1^3$ |         |         | A       |         |         | C       | S       |         |
| $B_1^3$ |         |         |         | A       | C       |         |         | S       |

→

|        | $A_1^0$ | $B_1^1$ | $A_1^2$ | $B_1^3$ | $B_1^0$ | $A_1^1$ | $B_1^2$ | $A_1^3$ |
|--------|---------|---------|---------|---------|---------|---------|---------|---------|
| $A_1^0$ | S       | C       | A       |         |         |         |         |         |
| $B_1^1$ | C       | SA      | C       | A       |         |         |         |         |
| $A_1^2$ | A       | C       | S       | C       |         |         |         |         |
| $B_1^3$ |         | A       | C       | S       |         |         |         |         |
| $B_1^0$ |         |         |         |         | S       | C       | A       |         |
| $A_1^1$ |         |         |         |         | C       | SA      | C       | A       |
| $B_1^2$ |         |         |         |         | A       | C       | S       | C       |
| $A_1^3$ |         |         |         |         |         | A       | C       | S       |

Figure A1. Left-hand side: vibrational symmetry expanded version of Hamiltonian matrix structure presented in Figure 1 (prolate top, z-axis *in plane*) of the main text for $J = 3$. Rows/columns are labelled by vibrational symmetry ($A_1$ or $B_1$) and $\Lambda$ (superscript). Right-hand side: a possible rearrangement of rows and columns that leads to separation of the overall Hamiltonian into 2 independent blocks.

Each submatrix uses only one symmetry of $f_m$ in a given $\Lambda$-block. The symmetry of $f_m$ alternates between successive $\Lambda$-blocks and starts with $f_m^+$ in one submatrix and $f_m^-$ in the other one. Thus, the $\pm$ sign in Eq. (A7) can be expressed through the value of $\Lambda$ and the value of starting symmetry in the $\Lambda = 0$ block as:

$$\langle f_m | \frac{d}{d\varphi} | f_{m'} \rangle = (-1)^{\Lambda+s} m \delta_{mm'}, \tag{A8}$$

where $s$ is the symmetry of the $\Lambda = 0$ block, defined as:

$$s = \begin{cases} 0 \text{ for } f_m^+ \text{ in } \Lambda = 0, \\ 1 \text{ for } f_m^- \text{ in } \Lambda = 0. \end{cases} \tag{A9}$$

The integral $\int_a^b f_m^\pm f_{m'}^\pm \, d\varphi$, can be calculated analytically for arbitrary limits $a$ and $b$. In the case of $\langle f_m | \hat{P}_{sym} | f_{m'} \rangle$ in Eq. (42) of the main text, $a = 2\pi/3$ and $b = 4\pi/3$, which results in the following solutions:

if $m = m' = 0$, then:

$$\langle f_m | \hat{P}_{sym} | f_{m'} \rangle = 1/3. \tag{A10}$$

if $m = m' \neq 0$, then:

$$\langle f_m | \hat{P}_{sym} | f_{m'} \rangle = \frac{1}{2\pi} \left( \frac{2\pi}{3} - (-1)^{\Lambda+s} \frac{\sin\left(\frac{4\pi}{3} m\right) - \sin\left(\frac{8\pi}{3} m\right)}{2m} \right), \tag{A11}$$

Finally, if $m \neq m'$, then:

$$\langle f_m | \hat{P}_{sym} | f_{m'} \rangle = \frac{1}{2\pi\sqrt{(\delta_{m0}+1)(\delta_{m'0}+1)}}$$

$$\times \left( \frac{\sin\left(\frac{4\pi}{3}(m-m')\right) - \sin\left(\frac{2\pi}{3}(m-m')\right)}{m-m'} \right.$$

$$+ \left. \frac{(-1)^{\Lambda+s} \sin\left(\frac{4\pi}{3}(m+m')\right) - (-1)^{\Lambda+s} \sin\left(\frac{2\pi}{3}(m+m')\right)}{m+m'} \right). \tag{A12}$$

**APPENDIX B. ROTATIONAL MATRICES U AND W**

The analytical expressions for matrices $U_{\Lambda\Lambda'}$ and $W_{\Lambda\Lambda'}$, introduced in Eqs. (20) and (21) for the asymmetric-top rotor terms and the Coriolis couplings are given by:[35]

$$U_{\Lambda\Lambda'} = \frac{1}{\sqrt{(1+\delta_{\Lambda 0})(1+\delta_{\Lambda' 0})}} \big(\lambda_+(J,\Lambda)\lambda_+(J,\Lambda+1)\delta_{\Lambda,\Lambda'-2}$$
$$+ \lambda_+(J,\Lambda')\lambda_+(J,\Lambda'+1)\delta_{\Lambda,\Lambda'+2} + (-1)^{J+\Lambda+p}\lambda_+(J,\Lambda'-1)\lambda_+(J,\Lambda'-2)\delta_{\Lambda,2-\Lambda'}\big), \quad \text{(B1)}$$

$$W_{\Lambda\Lambda'} = \frac{1}{\sqrt{(1+\delta_{\Lambda 0})(1+\delta_{\Lambda' 0})}} \big(\lambda_+(J,\Lambda)\delta_{\Lambda,\Lambda'-1} - \lambda_+(J,\Lambda')\delta_{\Lambda,\Lambda'+1}$$
$$+ (-1)^{J+\Lambda+p}\lambda_+(J,\Lambda'-1)\delta_{\Lambda,1-\Lambda'}\big), \quad \text{(B2)}$$

where

$$\lambda_\pm(J,\Lambda) = \sqrt{(J\pm\Lambda+1)(J\mp\Lambda)}. \quad \text{(B3)}$$